\documentclass[aps,pra,a4paper,twocolumn,showpacs,superscriptaddress,floatfix]{revtex4-2} 
\usepackage[utf8]{inputenc}
\usepackage{mathtools}
\usepackage{amsmath}
\usepackage{amsfonts}
\usepackage{graphicx}
\usepackage{longtable}
\usepackage{bbm}
 \usepackage{booktabs}
\usepackage{dsfont}
\usepackage{placeins}
\newcommand{\be}{\begin{equation}}
\newcommand{\ee}{\end{equation}}
\newcommand{\bea}{\begin{eqnarray}}
\newcommand{\eea}{\end{eqnarray}}

\newcommand{\ba}[1]{\left(\begin{array}{#1}}
\newcommand{\ea}{\end{array}\right)} 
\usepackage{multirow}
\usepackage{hhline}
\usepackage{xcolor}

\usepackage{float}
\usepackage{placeins}

\begin{document}
\title{Dicke superposition probes for noise-resilient \\ Heisenberg and super-Heisenberg metrology}
\author{Sudha}
\email{tthdrs@gmail.com}
\affiliation{Department of Physics, Kuvempu University,
Shankaraghatta--577 451, Karnataka, India}
\affiliation{Inspire Institute Inc., Alexandria, Virginia 22303, USA}
\author{B.\ N.\ Karthik}
\affiliation{Department of Physics, Kuvempu University,
Shankaraghatta--577 451, Karnataka, India}
\author{K.\ S.\ Akhilesh}
\affiliation{School of Computer Science and Engineering,
RV University, Bangalore--560 059, Karnataka, India}
\author{A.\ R.\ Usha Devi}
\affiliation{Department of Physics, Bangalore University,
Bangalore--560 056, Karnataka, India}
\affiliation{Inspire Institute Inc., Alexandria, Virginia 22303, USA}

\date{\today}
\begin{abstract}
Phase sensing with entangled multi-qubit states in the presence of noise is a central theme of modern quantum metrology. The present work investigates Dicke state superposition probes for quantum phase sensing under parameter encoding generated by one- and two-body interaction Hamiltonians.  Under linear collective-spin encoding, near-optimal Dicke superposition states are shown to exhibit significantly enhanced robustness against phase damping noise compared with Greenberger--Horne--Zeilinger (GHZ), W-superposition, and balanced Dicke states, while maintaining favorable metrological performance under realistic decoherence channels.
For two-body interactions, optimal probe states maximizing the quantum Fisher information are identified. Their noise resilience and metrological scaling behaviour under phase damping, amplitude damping, and global depolarizing channels are analyzed.  The associated near-optimal Dicke superposition states are found to exhibit improved resilience to phase damping, for the system sizes considered.  These results establish tailored near-optimal Dicke state superposition probes as versatile and noise-resilient resources for Heisenberg and super-Heisenberg quantum phase sensing governed by one- and two-body interactions.
\end{abstract}
\maketitle 
\section{Introduction}  
Quantum metrology exploits non-classical correlations to enhance the precision of parameter estimation beyond the standard quantum limit (SQL) ~\cite{bc,bollinger,huelga,sethlyod,sym1a,prlmaccone06,mparis,pezze,Reilly2023}. Entangled multi-particle probe states and parameter-encoding Hamiltonian interactions play a key role in attaining quantum advantages in interferometry, spectroscopy and sensing platforms with atoms, photons and solid-state qubits~\cite{pezze,Reilly2023,hylus,pral25}.  A central question concerns the identification of multipartite entangled probes and parameter-encoding Hamiltonians that remain metrologically useful in noisy scenarios~\cite{pezze,pral25,Frowis2014}.  Recent literature has assessed quantum advantage in metrology by comparing the quantum Fisher information (QFI) of entangled probe states with separability bounds tailored to specific Hamiltonians, thereby identifying protocols that genuinely outperform all separable strategies~\cite{pezze,hylus,ujwal25,pral25}. In particular, Imai \emph{et al.}~\cite{pral25} analyzed nonlinear Hamiltonians and derived general QFI-based criteria for metrologically useful entanglement beyond the case of non-interacting Hamiltonians, while Bhattacharyya \emph{et al.}~\cite{ujwal25} investigated quantum sensing of even- versus odd-body interactions, demonstrating how nonlinear encoding  Hamiltonians  lead to distinct scaling behaviors and entanglement requirements~\cite{pral25,ujwal25}.  

Most studies of noisy quantum phase sensing have  focused on a restricted set of canonical probe families, such as Greenberger--Horne--Zeilinger (GHZ) states~\cite{huelga,Escher2011,SciRep2017}, $W$ superpositions~\cite{NJP2014,wwb23}, and Dicke states~\cite{toth,TothDickeQFI2017, pra2024}, whose QFI scaling and noise robustness have been systematically studied. Beyond these standard probes, there is increasing interest in engineered superpositions of Dicke states~\cite{Yin2011,DS2019, markham2021,ChenPRL2023}. From a foundational standpoint, such superpositions exhibit nontrivial multipartite quantum correlations, quantified for example, by geometric measures of quantum discord~\cite{Yin2011} that can persist even in regimes where entanglement is strongly suppressed by noise.  From an experimental perspective, recent demonstrations of on-chip generation and coherent control of superpositions spanning the entire Dicke manifold~\cite{ChenPRL2023}, together with experimentally feasible protocols for generating metrologically useful Dicke superposition states through rapid adiabatic passage (RAP)~\cite{Carrasco2024Dicke} and cavity-mediated state engineering~\cite{Sawant2023}, establish these states as experimentally accessible multipartite resources in state-of-the-art quantum platforms.

In this paper, Dicke-state superposition probes are investigated for quantum phase sensing generated by one- and two-body qubit interaction Hamiltonians, with particular emphasis on the scaling of the QFI and robustness against dephasing, amplitude damping, and global depolarizing noise. 

The present work extends earlier studies in two important directions. 
First, the metrological performance of Dicke superposition probes exhibiting near-Heisenberg scaling under linear collective-spin encoding is investigated under realistic noise models through a systematic comparative analysis with GHZ, W-superposition, and balanced Dicke states. Second, for a representative class of collective two-body interaction Hamiltonians, the \emph{optimal} probe states are identified analytically.  Their noise resilience, and scaling behaviour are subsequently analyzed. In addition, the corresponding Dicke superposition states with quantum Fisher information closest to the optimal value are identified and their noise resilience is benchmarked against that of the optimal probe states.

The organization of the paper is as follows. Section~II reviews the basic framework of parameter estimation in quantum metrology. Tailored Dicke superposition states exhibiting near-Heisenberg scaling of the QFI for one-body interaction Hamiltonians are presented in Section~III, and their noise robustness is compared with that of GHZ, W-superposition, and balanced Dicke probes. Section~IV extends the analysis to  two-body interaction Hamiltonians, identifying optimal and near-optimal Dicke superposition probes and examining their noise resilience. Section~V gives a summary and concluding remarks. 
 
\section{Parameter estimation and  Fisher information}
Parameter estimation is a core task underlying precision measurements in both classical and quantum metrology. Depending on the application, the parameter of interest may correspond to a frequency, phase, magnetic field, temperature, or other physical quantity to be estimated. Despite this diversity, any estimation protocol can be conceptually decomposed into three essential stages: (i) preparation of input probe, (ii) parameter encoding, and  (iii) probe readout via measurement and statistical inference.

\subsection{Classical Cram\'er-Rao bound }
In classical parameter estimation, imperfections in probe preparation, system control, and measurement accuracy limit the achievable precision. For any unbiased estimator $\hat\theta$ of a parameter $\theta$, the ultimate precision $\Delta{\theta}$ achievable in a classical parameter-estimation scheme is bounded from below by the Cram\'er-Rao bound~\cite{Helstrom1976,Kay1993} 
\be 
(\Delta {\theta})^2 \geq \frac{1}{F_C(\theta)} 
\ee 
where 
\be 
\label{clfi} F_C(\theta)= \sum_x\, p(x\vert\theta) \left(\frac{\partial\ln p(x\vert\theta)}{\partial\,\theta}\right)^2 
\ee 
is the classical Fisher information and $p(x\vert\theta)$ denotes the probability of obtaining the outcome $x$ for the random variable $X$ used to estimate  the parameter $\theta$. The Cram\'er-Rao bound sets the fundamental lower limit on the variance $(\Delta \theta)^2$ of any unbiased estimator $\hat{\theta}$ of a parameter $\theta$ in classical estimation theory. 

\subsection{Quantum parameter estimation} 
In the quantum scenario, the measurement precision is bounded by~\cite{Helstrom1976}
\be 
\label{qcrb} 
(\Delta{\theta})^2 \geq \frac{1}{F_Q(\hat{\rho},\hat{H})} \ee 
where $F_Q(\hat{\rho},\hat{H})$ denotes the QFI,  $\hat\rho$ corresponds to the probe state and $\hat{H}$ is the  Hamiltonian generating the  unitary encoding   $\hat{U}(\theta)=e^{-i\hat{H}\theta}$.  After encoding, the probe state transforms to $\hat{\rho}_\theta=\hat{U}(\theta)\,\hat{\rho}\, 
\hat{U}^\dagger(\theta).$  Performing a generalized measurement described by a positive-operator-valued measure (POVM) $\hat{E}\equiv\left\{\hat{E}_x, \ \displaystyle\sum_x \hat{E}_x=I\right\}$ on the encoded state $\hat{\rho}_\theta$ yields classical Fisher information
 \be 
 \label{qfi} 
 F_C(\theta)= \sum_x\, \frac{1}{p(x|\theta)} \left(\frac{\partial p(x\vert\theta)}{\partial\theta}\right)^2 
 \ee 
 where $$p(x|\theta)=\mbox{Tr}\,(\hat{E}_x\,\hat{\rho}_\theta).$$ The QFI $F_Q(\hat{\rho},\hat{H})$ is defined as the maximum of the classical Fisher information over all possible POVMs and sets the ultimate precision allowed by the quantum Cramér–Rao bound~\cite{bc}:
 \be 
 \label{qfimax} 
 F_Q(\hat{\rho},\hat{H})=\underset{\hat{E}}{\rm max}\, F_C(\theta). 
 \ee 
 The QFI can, in principle, be attained by projective measurements in the eigenbasis of the symmetric logarithmic derivative (SLD) operator~\cite{Helstrom1976,bc} 
 \be \label{sld} 
 \frac{\partial{\hat{\rho}_\theta}}{\partial{\theta}}=\frac{1}{2}\,\left(L(\hat{\rho}_\theta)\hat{\rho}_\theta+ \hat{\rho}_\theta\,L(\hat{\rho}_\theta)\right)
 \ee 
 in terms of which it can be expressed as 
 \be 
 \label{qfisld} 
F_Q(\hat{\rho},\hat{H})=\mbox{Tr}\,\left[\hat{\rho}_\theta\,L^2(\hat{\rho}_\theta)\right].  
 \ee 
The quantum Cramér–Rao bound (\ref{qcrb}) thus implies that, for a given metrological protocol specified by the probe state and the parameter-encoding Hamiltonian, the QFI  $F_Q(\hat{\rho},\hat{H})$ quantifies the ultimate sensitivity of the probe state to minute changes in the parameter  $\theta$ being estimated. 

For an arbitrary  mixed probe state 
$\hat{\rho} = \displaystyle\sum_i \lambda_i |\psi_i\rangle\langle\psi_i|$ 
under unitary encoding generated by a Hamiltonian $\hat H$, the QFI is given by~\cite{bc}
\begin{equation}
\label{qf1}
F_Q(\hat{\rho},\hat{H})
=2 \sum_{i,j}
\frac{(\lambda_i-\lambda_j)^2}{\lambda_i+\lambda_j}
\left|
\langle \psi_i | \hat{H} | \psi_j \rangle
\right|^2 ,
\end{equation}
where the sum runs over all indices satisfying $\lambda_i+\lambda_j\neq0$.  In the case of  a pure probe state $\hat{\rho}=|\psi\rangle\langle\psi|$, the QFI  reduces to
\begin{equation}
\label{qfp}
F_Q(|\psi\rangle,\hat{H})
=
4\,\Delta^2 \hat{H}
=
4\left(
\langle \psi | \hat H^2 | \psi\rangle
-
\langle \psi | \hat H | \psi\rangle^2
\right),
\end{equation}
where $\Delta^2 \hat H$ denotes the variance of $\hat H$ in the state $|\psi\rangle$.

For unitary parameter encoding  on pure probe states, the maximal QFI achievable for a given Hamiltonian $\hat H$ is given by~\cite{PangBrun2014} 
\begin{equation}
\label{defqfi}
F_Q^{\rm max}(|\psi\rangle_{\rm opt}, \hat H)
= 4\,\Delta^2 \hat{H} =
(\lambda_{\rm max}-\lambda_{\rm min})^2,
\end{equation}
where $\lambda_{\rm max}$ and $\lambda_{\rm min}$ denote the largest and smallest eigenvalues of $\hat H$, respectively. This bound is attained by the probe state
\begin{equation}
\label{defopt}
|\psi\rangle_{\rm opt}=
\frac{|\phi_{\rm max}\rangle + |\phi_{\rm min}\rangle}{\sqrt{2}},
\end{equation}
where $|\phi_{\rm max}\rangle$ and $|\phi_{\rm min}\rangle$ are the eigenstates of the Hamiltonian $\hat H$ corresponding to the eigenvalues  $\lambda_{\rm max}$ and $\lambda_{\rm min}$, respectively. For any other choice of pure probe state, the achievable QFI is given by ~(\ref{qfp}), while for mixed probe states it is given by Eq.~(\ref{qf1}). 

As an important example, consider the collective-spin Hamiltonian
\be
\label{csh}
\hat{H} = \hat{\mathbf{J}}\cdot \hat{\boldsymbol{n}}\equiv \hat{\mathbf{J}}_{\boldsymbol{n}}
\ee
where $\hat{\mathbf{J}}=(\hat{J}_x,\hat{J}_y,\hat{J}_z)$ denotes the collective-spin
operator of an $N$-qubit system, with  $\hat{J}_\alpha=\frac{1}{2}\,\sum_i\,\hat{\sigma}_{i\alpha}$, $\alpha=x,y,z$;  $\hat{\sigma}_{i\alpha}$ denoting the Pauli matrices of the $i^{\rm th}$ qubit, and $\hat{\boldsymbol{n}}$ denotes a unit vector specifying the direction. 
The extremal eigenvalues of the Hamiltonian in (\ref{csh}) are given by $\lambda_{\rm max}=N/2$ and $\lambda_{\rm min}=-N/2$, yielding  
\begin{equation}
F_Q = N^2
\end{equation}
with the corresponding probe state attaining this bound being the
$N$-qubit Greenberger--Horne--Zeilinger (GHZ) state~\cite{ghz},
\be
|\mathrm{GHZ}\rangle = \frac{1}{\sqrt{2}}\, \left(|0\rangle^{\otimes N} + |1\rangle^{\otimes N}\right).
\ee

In the following, quantum metrology protocols employing multiqubit probe states and single-parameter unitary encodings generated by one- and two-body interaction Hamiltonians will be discussed.  For a given encoding Hamiltonian and fixed system size $N$,
a probe state that maximizes the QFI is referred to as an \emph{optimal} state~\cite{Frowis2014,Volkoff2016}. 
Dicke superposition states whose QFI lies closest to the  corresponding optimal value will be referred to as \emph{near-optimal} probe states. The identification of such states is of practical interest, as they retain the essential metrological advantage of the optimal probe. The metrological performance of optimal and near-optimal Dicke superposition probes under realistic noise channels is investigated in the following sections.

\subsection{Encoding with linear and nonlinear interactions}

A wide range of metrological strategies for achieving high-precision measurements have been explored in the literature~\cite{pezze}. These approaches are commonly
classified according to the nature of the Hamiltonian that encodes the parameter onto an $N$-qubit probe state. In linear (non-interacting or one-body interaction) quantum metrology~\cite{Helstrom1976,bc,prlmaccone06,pezze}, the
encoding multi-qubit Hamiltonian is given by ~(\ref{csh}).  
This  Hamiltonian is 
linear in the probe constituent qubits and  does not generate any entanglement 
during the encoding process. Collective, non-interacting generators of this form arise naturally in standard phase-estimation protocols and therefore provide a natural testbed for assessing the metrological utility of Dicke superposition probes. 

In a parameter estimation protocol involving a linear parameter-encoding Hamiltonian and  $N$-qubit separable probe states,  the QFI scales as~\cite{pezze}
\begin{equation}
F_Q(\hat{\rho}_{\rm sep},\hat{H}) \propto N,
\end{equation}
and correspondingly leads to the SQL or shot-noise limit (SNL) precision: 
\begin{equation}
\Delta \theta \sim \frac{1}{\sqrt{N}}.
\end{equation}
In contrast, genuinely entangled $N$-qubit  probe states can achieve Heisenberg scaling~\cite{Kok2012,pezze}
of QFI under linear Hamiltonian encoding, with
$F_Q(\hat{\rho}_{\rm ent},\hat{H}) \propto N^2$
in which case enhanced precision of estimation, the Heisenberg limit (HL) is achieved:  
\begin{equation}
\Delta \theta \sim \frac{1}{N}.
\end{equation}

Nonlinear metrology, on the other hand, employs parameter-encoding Hamiltonians containing  many-body interaction 
terms~\cite{Luis2004,prl07,shaji08,Boixo2008,Boixo2009,Pezze2009,nature10,Zwierz2010,2010PRLEntNotCritical,rmp18,pra14r,pra17usa,natphys11,TsarevEtAl,njp10,prx18ph,Deng2021,Lordi2025NoiseConstraints,prrs2022,natphys22,Chu2023,pral25,ujwal25,2017DMInt,njp08caves,
sewell14,sb18,2013sewelColdAtomExpt,Ran2021,prl22}. Typically, the collective-spin Hamiltonians with $k$-body interactions of the form 
\begin{eqnarray}
\hat{H}_k &=& \left(\hat{\mathbf{J}}_{\boldsymbol{n}}\right)^k
=\frac{1}{2^k}\sum_{i_1,\dots,i_k=1}^N \hat{\sigma}_{\boldsymbol{n}}^{(i_1)}\cdots\hat{\sigma}_{\boldsymbol{n}}^{(i_k)},
\label{Hk_expanded}
\end{eqnarray} 
where the sum in (\ref{Hk_expanded}) includes $k$-body interaction terms with $i_1<\cdots<i_k,\  k=1,2,\ldots $ have been studied~\cite{pral25, ujwal25}. For $k>1$, $\hat{H}_k$ contains nonlinear $k$-body interactions generating the encoding unitary operation. 
Such encoding unitaries can dynamically generate quantum correlations, thereby allowing even initially separable probe states to surpass the SNL~\cite{Luis2004,Boixo2008,2010PRLEntNotCritical,rmp18}. 
It has been shown that the Hamiltonians containing $k$-body interaction terms can, in idealized settings, enable so-called super-Heisenberg precision scaling
$\Delta {\theta}\sim 1/N^{k}$~\cite{prl07,shaji08, Boixo2008}. 

The metrological advantages predicted by both linear and nonlinear schemes are often severely degraded once realistic noise is taken into account~\cite{ncnoisy2012,pl1,prl14Maccone,prl1122014,dicken1,markham2021,pra2024}. A paradigmatic example is the $N$-qubit  GHZ state, which attains Heisenberg-limited (HL) scaling for linear generators in the noiseless limit. However, GHZ states are extremely fragile to noise, and their metrological advantage diminishes quickly under decoherence.  Analogous situations arise in nonlinear metrology: Although nonlinear Hamiltonians can, in principle, enable super-Heisenberg scaling in idealized metrological scenarios, realistic noise typically suppresses the achievable precision well below the ideal scaling of QFI. These limitations, imposed by inevitable noise, are not specific to any single class of probe states; rather, they reflect the general fragility of metrological advantages, linear or nonlinear, in realistic experimental conditions. Thus, a systematic evaluation of metrological performance under explicit noise models is crucial for identifying practically robust probes and sensing protocols. Accordingly, identifying probe states that combine better phase sensitivity with substantial noise resilience remains a key issue in quantum metrology.

The generators of the form (\ref{Hk_expanded}) preserve permutation symmetry and therefore confine the unitary parameter-encoding strategy to the symmetric subspace. This motivates a detailed investigation on $N$-qubit permutation-symmetric Dicke superposition probes for noise-resilient quantum metrology with one- and two-body interaction Hamiltonians.

\section{Dicke superposition probes for noise-resilient linear metrology}

Multiqubit states that are invariant under interchange of qubits are an extensively studied class of quantum states, notable both for their experimental relevance and elegant mathematical structure~\cite{Majorana1932,Dicke1954, Arecchi1972,KitagawaUeda1993,Bastin2009,Aulbach2010,Markham2011,UshaDevi2011,Lucke2011,SchleierSmith2010,Bohnet2016,Zeiher2016,Baguette2014,WangMarkham2012,Wiesniak2021,MunozArias2023,revsymmetric2025}.  Prominent examples include GHZ~\cite{ghz}, W~\cite{DurVidalCirac2000}, and Dicke states~\cite{Dicke1954}. States obeying symmetry under interchange of $N$-constituent qubits confine themselves to the $(N+1)$-dimensional subspace spanned by Dicke states
\begin{eqnarray}
|D_{N-l,l}\rangle &\equiv& \left| J = \frac{N}{2}, M = \frac{N}{2}-l \right\rangle, 
\quad l = 0,1,\ldots,N, \nonumber \\
&=& \sqrt{\frac{l!(N-l)!}{N!}} 
\displaystyle\sum_{\mathcal{P}}
\mathcal{P}\!\left(|0\rangle^{\otimes N-l}\otimes \vert 1\rangle^{\otimes l}\right) 
,\end{eqnarray}
where ${\cal P}$ runs over all distinct permutations of the qubits; this set $\{|D_{N-l,l}\rangle\}$ of $N+1$ basis states is a simultaneous eigenbasis of the collective-spin operators $\hat{J}^2~=~\hat{J}_x^2~+~\hat{J}_y^2~+~\hat{J}_z^2$ and $\hat{J}_z$: 
\begin{eqnarray}
\hat{J}^2\,|D_{N-l,l}\rangle&=&\frac{N}{2}\left(\frac{N}{2}+1\right)\,|D_{N-l,l}\rangle \nonumber \\ 
\hat{J}_z|D_{N-l,l}\rangle&=& \left(\frac{N}{2} - l \right)\,|D_{N-l,l}\rangle. 
\end{eqnarray}
Recent experiments have demonstrated remarkable control over multiqubit Dicke states across diverse physical platforms, including photonic systems, cold atoms, and trapped ions~\cite{ChenPRL2023,exptdicke3,exptdicke4,exptdicke5,exptdicke6,pnas10k,pra21bmterhal}. State-of-the-art experiments have enabled on-chip generation and collectively coherent control of the entire family of four-photon Dicke states, with the underlying architecture and control scheme extendable to larger photon numbers~\cite{ChenPRL2023}. Complementing these experimental advances, recent proposals based on rapid adiabatic passage (RAP) and one-axis twisting (OAT) dynamics~\cite{KitagawaUeda1993,Carrasco2024Dicke} as well as cavity-mediated state engineering~\cite{Sawant2023} have demonstrated feasible routes for generating Dicke  and 
Dicke-state superpositions with significant metrological gain.

\subsection{Linear metrology with Dicke superpositions}  
\label{subsec:A}

For the Dicke state $\vert D_{N-l, l}\rangle$, the QFI associated with unitary parameter encoding generated by the linear Hamiltonian $\hat{H}_{k=1}=\hat{\mathbf J}_{\boldsymbol n}$ (see Eq.~(\ref{csh})) 
is given by
\begin{eqnarray}
&&F_Q(\vert D_{N-l, l}\rangle,\hat{H}_1)
=4\,\max_{\boldsymbol n}\,(\Delta \hat{\mathbf J}_{\boldsymbol n})^2 \nonumber \\
&&\ \ \ \ \ \ = N + 2\,l\,(N-l),
\ \ l=0,1,2,\ldots,N . 
\end{eqnarray}
The optimal encoding direction lies in the $XY$ plane. The QFI exhibits a quadratic dependence on the excitation number $l$ and is maximized for  $l=N/2$ i.e., $F_Q(\vert D_{N/2, N/2}\rangle,\hat{H}_1)=N(N+2)/2$ for even $N$. 
The resulting QFI scaling surpasses the SNL while remaining below the Heisenberg limit.

To assess whether this limitation can be overcome, superpositions of two distinct Dicke states
are considered:
\begin{eqnarray}
\label{dsdefn}
\vert D^{(N)}_{l,l'}\rangle
&=&
\frac{1}{\sqrt{2}}
\left(
\vert D_{N-l, l}\rangle
+
\vert D_{N-l', l'}\rangle
\right), \\ 
&& \qquad l\neq l';\ l,l'=1,2,\ldots, N. \nonumber
\end{eqnarray}
For Dicke superposition probes, the QFI 
$F_Q(\vert D^{(N)}_{l, l'}\rangle,\hat{H}_1)=4\,\max_{\boldsymbol n}\,(\Delta \hat{\mathbf J}_{\boldsymbol n})^2$ for
certain choices of $(l,l')$ yields a quadratic scaling  with  $N$. More specifically, 
 the QFI for odd-$N$ superpositions with
\begin{equation}
\label{odd}
\frac{N}{2}-l=\mp \frac{1}{2},\ \ 
\frac{N}{2}-l'=\pm\frac{3}{2}
\end{equation} 
is evaluated to be 
\begin{equation}
\label{sdo}
F_Q(\vert D^{(N)}_{l, l'}\rangle,\hat{H}_1)
\approx
8.4641+\frac{3}{4}(N-3)(N+5).
\end{equation}
For even $N$ Dicke superpositions $\vert D^{(N)}_{l, l'}\rangle$  with 
\begin{equation}
\label{even}
\frac{N}{2}-l=\pm 1,
\ \ 
\frac{N}{2}~-~l'=\mp 1,
\end{equation} 
the QFI is given by 
\begin{equation}
\label{sde}
F_Q(\vert D^{(N)}_{l, l'}\rangle,\hat{H}_1)
=
4+\frac{3}{4}(N-2)(N+4).
\end{equation}
Representative QFI values of near-optimal Dicke superposition probes for system sizes ranging from $N=3$ to $N=20$ are given in Table~\ref{tab:dicke_superposition_qfi}, illustrating the near-Heisenberg $\mathcal{O}(N^2)$ scaling.

\begin{table}[h]
\caption{QFI  
$F_Q(|D^{(N)}_{l,l'}\rangle,\hat{H}_1)
~=~ 4\max_{\boldsymbol n}(\Delta\hat{\mathbf J}_{\boldsymbol n})^2$
for near-optimal Dicke superposition probes 
$|D^{(N)}_{l,l'}\rangle = (|D_{N-l,l}\rangle + |D_{N-l',l'}\rangle)/\sqrt{2}$
 for system sizes ranging from $N=3$ to $N=20$. For each $N$, the listed pairs $(l,l')$ maximize the QFI
under linear collective-spin encoding generated by $\hat{H}_1 = \hat{\mathbf J}_{\boldsymbol{n}}$.
The listed values illustrate the near-Heisenberg
$\mathcal{O}(N^2)$ scaling of these probes.}
\label{tab:dicke_superposition_qfi}
\begin{ruledtabular}
\begin{tabular}{ccc}
$N$ & $(l,l')$ & $F_Q(|D^{(N)}_{l,l'}\rangle,\hat{H}_1)$ \\
\hline
3 & $(0,\,2),\ (1,\,3)$ \  & 8.46 \\
4 & $(3,\,1)$ & 16 \\
5 & $(1,\,3),\ (2,\, 4)$  & 23.48 \\
7 & $(2,\,4),\ (3,\,5)\ $ & 44.49 \\
15 & $(6,\,8),\ (7,\,9)$ & 188.46  \\ 
20 & $(11,\,9)$ & 328 \\ 
\end{tabular}
\end{ruledtabular}
\end{table}
 To place these results in context, it is instructive to compare them with other prominent multipartite probe states, in particular with the $N$-qubit W--superposition state
$\lvert \mathrm{W\bar W}\rangle ~\equiv~ \lvert D^{(N)}_{1,N-1}\rangle$,
corresponding to an equal superposition of the single-excitation Dicke
state and its spin-flipped counterpart. Under encoding by the
Hamiltonian  $\hat{H}_1=\hat J_z$, the QFI of $N$-qubit W superposition state is
$F_Q(\lvert D^{(N)}_{1,N-1}\rangle,\hat J_z) = (N-2)^2$ for $N \ge 6$~\cite{wwb23}
and it asymptotically approaches the Heisenberg-limit 
$F_Q(\lvert \mathrm{GHZ}\rangle,\hat J_z) = N^2$ attained by the GHZ state. 

\begin{figure}[ht]
	\centering
		\includegraphics[width=0.45\textwidth]{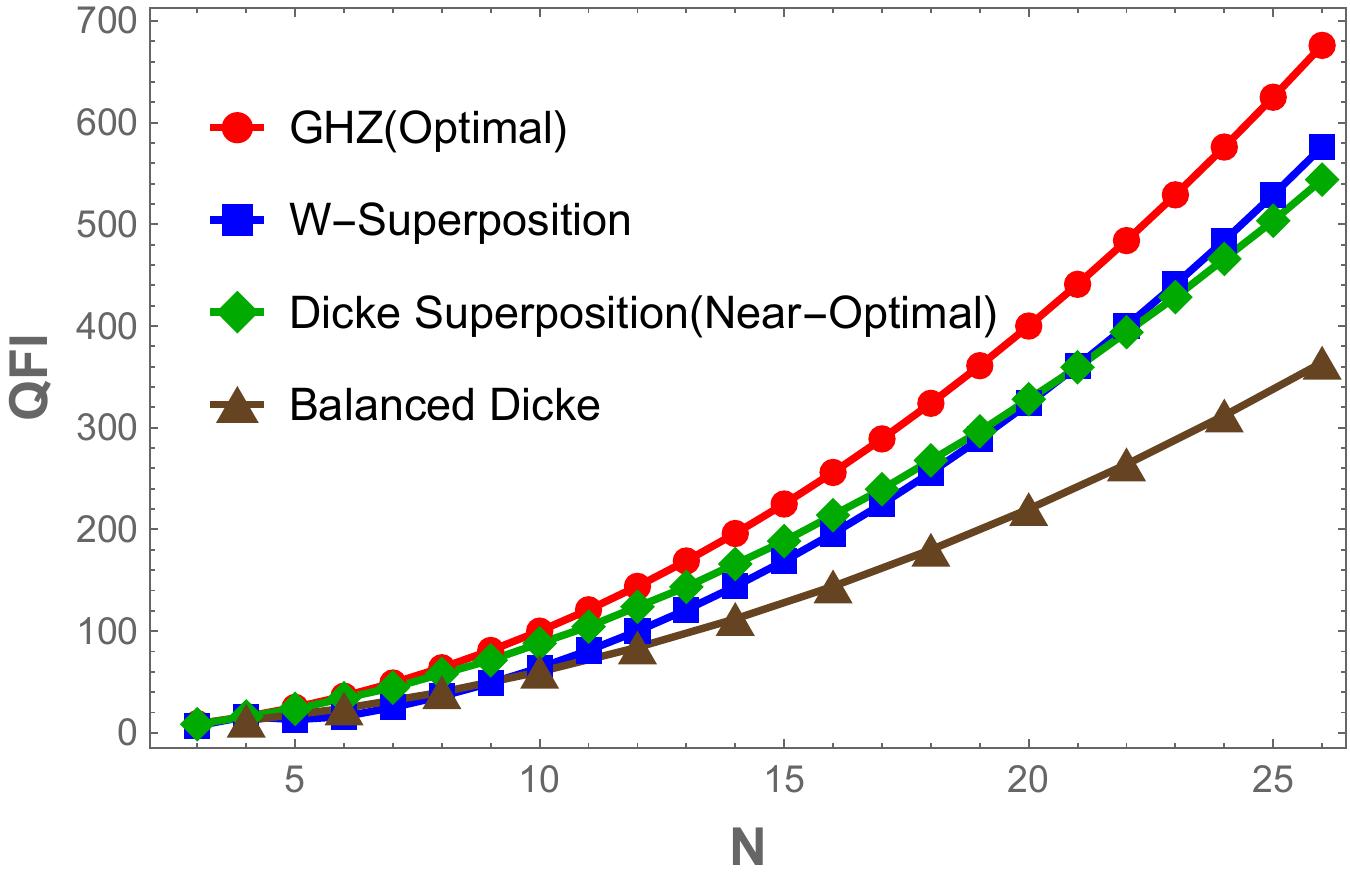}   
		\caption{QFI under collective-spin encoding $\hat J_{\boldsymbol n}$ for near-optimal
Dicke superposition states $\lvert D^{(N)}_{l,l'}\rangle$, compared with the
W--superposition state $\lvert D^{(N)}_{1,N-1}\rangle$, the balanced Dicke state
$\lvert D_{N/2,N/2}\rangle$ (for even $N$), and the GHZ state.
The Dicke superposition probes exhibit a modest but systematic enhancement in QFI relative to the
W--superposition state for system sizes $N\lesssim 20$. }	
\label{1}
\end{figure}

A comparison of the QFI of near-optimal Dicke superposition states $\lvert D^{(N)}_{l,l'}\rangle$ (see Eqs.~(\ref{sdo}), (\ref{sde}))  with GHZ, W-superposition, and balanced Dicke states $\lvert D_{N/2,N/2}\rangle$ (for even $N$)~\cite{pnas10k,BDBastin}, under collective-spin encoding by $\hat{H}_1=\hat J_{\boldsymbol n}$
is presented in Fig.~\ref{1}. 

Over the range of system sizes shown, the near-optimal Dicke superposition states  exhibit a systematic enhancement in QFI relative to the W--superposition
state holding up to $N\approx 20$. This identifies a regime in which near-optimal Dicke superposition probes
offer a systematic enhancement in QFI relative to the
W--superposition states.  {These observations motivate a detailed examination of the robustness of Dicke superposition probes under realistic noisy environments.}

\subsection{Noise resilience of near-optimal Dicke superposition states}
The noise resilience of near-optimal Dicke superposition probes 
under collective-spin encoding generated by $\hat J_{\boldsymbol{n}}$ is investigated below under three standard noise models: phase damping,  amplitude damping,   and global depolarization~\cite{nc}. These noise models capture dominant decoherence mechanisms
encountered in experimental platforms such as trapped ions~\cite{Ion}, Rydberg-atom arrays~\cite{Rydberg}, and cavity-mediated collective-spin systems~\cite{ColdAtoms}.

For local noise processes, the channel acts independently on each qubit, whereas global depolarization acts collectively on the full 
$N$-qubit state. A generic local noise channel is specified by single-qubit Kraus operators $\{\hat E_k\}$ satisfying $\sum_k \hat E_k^\dagger \hat E_k=\mathbb{I}_2$ and its action on an $N$-qubit state $\hat{\rho}$ is given by 
\begin{equation}
\label{outputn}
\hat\rho_{\mathrm{noisy}}=
\sum_{k_1,\ldots,k_N}
\left(\hat E_{k_1}\otimes\cdots\otimes\hat E_{k_N}\right)
\hat\rho
\left(\hat E^\dag_{k_1}\otimes\cdots\otimes\hat E^\dag_{k_N}\right).
\end{equation}
The Kraus operators for amplitude damping are given by
\begin{equation}
\hat E_1 = |0\rangle\langle 0| + \sqrt{1-p}\,|1\rangle\langle 1|,\qquad
\hat E_2 = \sqrt{p}\,|0\rangle\langle 1|.
\end{equation}
Phase damping is described by the Kraus operators
\begin{eqnarray}
\hat E_1&=&\sqrt{1-p}\,\mathbb{I}_2, \qquad \hat E_2=\sqrt{p}\,\lvert 0\rangle\langle 0\rvert, \nonumber \\
\hat E_3&=&\sqrt{p}\,\lvert 1\rangle\langle 1\rvert
\end{eqnarray} 
where $\mathbb{I}_2$ denotes $2\times 2$ identity matrix. 
Global depolarization acts collectively and transforms the state as 
\begin{equation}
\label{globaldep}
\hat\rho_{\mathrm{noisy}}
=
(1-p)\hat\rho+\frac{p}{2^N}\mathbb{I}_{2^N} ,
\end{equation}
For each noise model, the noisy output state $\hat\rho_{\mathrm{noisy}}$ is obtained
using Eq.~(\ref{outputn}) or Eq.~(\ref{globaldep}). The corresponding QFI is then computed using (\ref{qf1}), and the phase
sensitivity $\Delta\theta$ is evaluated using  the quantum Cramér--Rao bound,
(\ref{qcrb}).

\begin{figure}[t] 
 \centering 
 \includegraphics[width=0.45\textwidth]{fi12pd.png} \\[0.3cm]
 \includegraphics[width=0.45\textwidth]{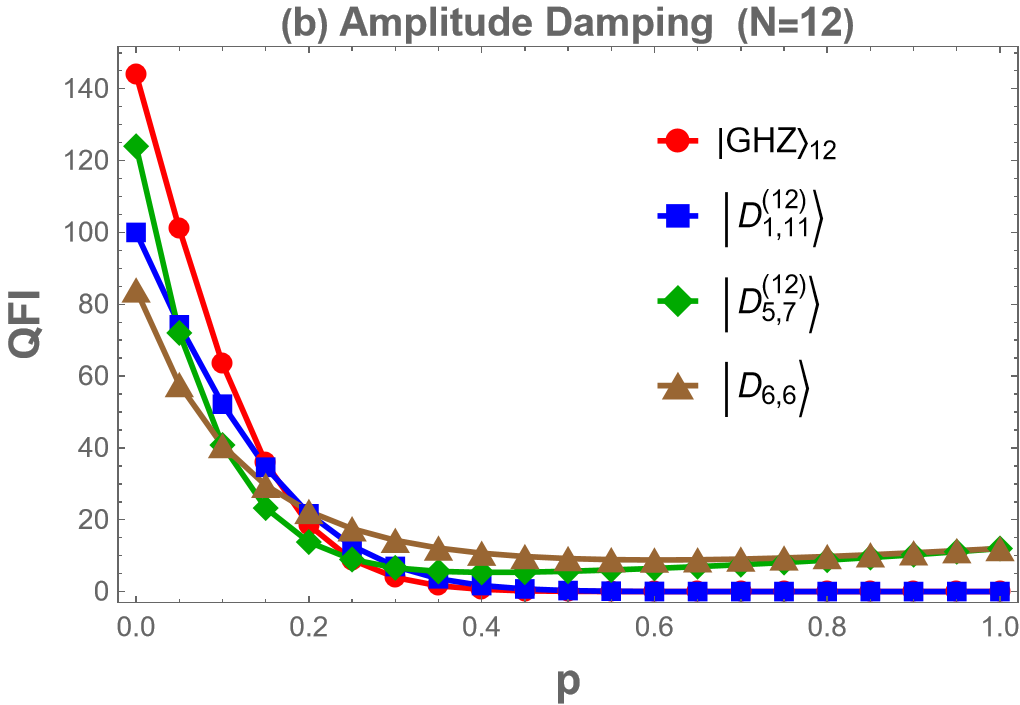} \\[0.3cm]
 \includegraphics[width=0.45\textwidth]{fi12dc.png} 
 \caption{QFI of multiqubit probe states as a function of the noise parameter $p$
under collective-spin encoding Hamiltonian $\hat{H}_1=\hat J_{\boldsymbol{n}}$. The near-optimal Dicke superposition state
$\vert D^{(12)}_{5,7}\rangle$ is compared with GHZ, W--superposition $\lvert D^{(12)}_{1,11}\rangle$, and
balanced Dicke state $\lvert D_{6,6}\rangle$ under (a) phase damping and (b) amplitude damping and  (c) global depolarization.} 
\label{fi8}
  \end{figure}

\begin{figure}[t]  
 \centering
  \includegraphics[width=0.45\textwidth]{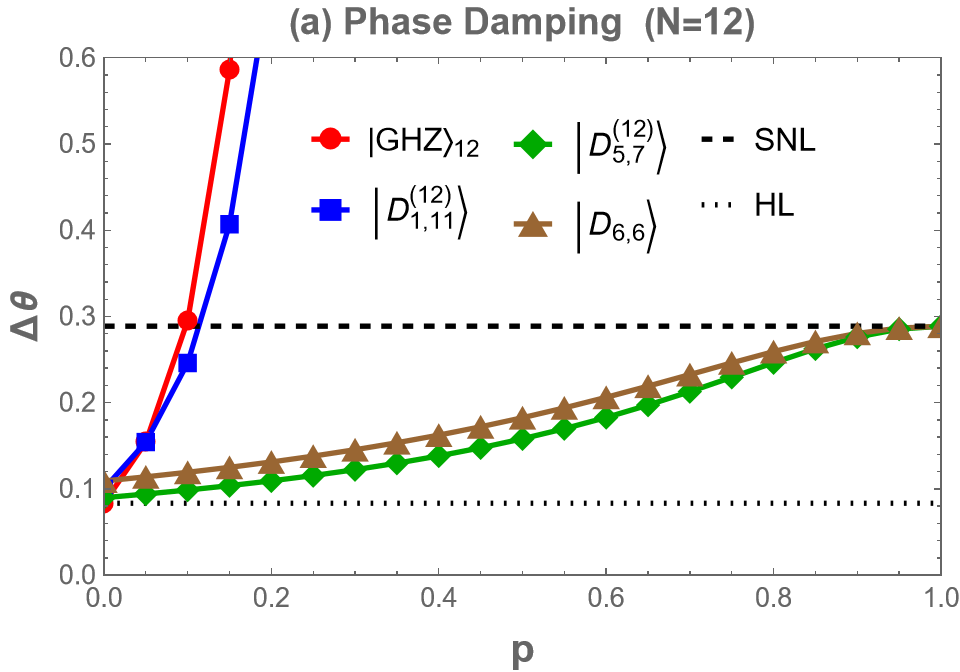} \\[0.3cm]
  \includegraphics[width=0.45\textwidth]{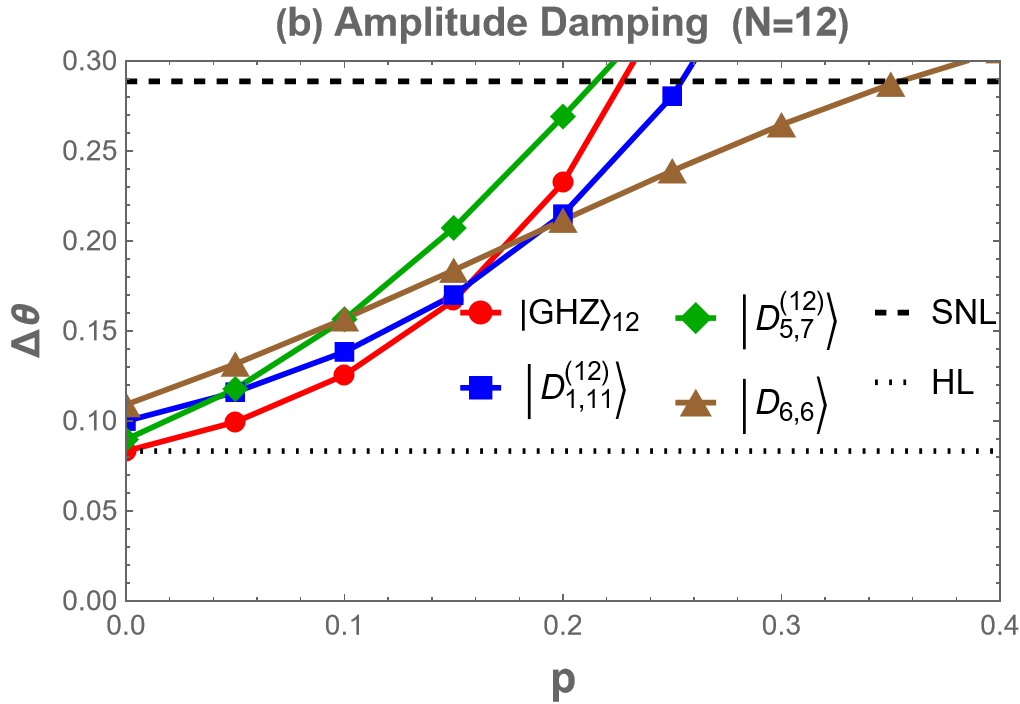} \\[0.3cm]
  \includegraphics[width=0.45\textwidth]{ps12dc.png} 
 \caption{Phase sensitivity $\Delta\theta$ of multiqubit probe states as a function of
the noise parameter $p$ under encoding Hamiltonian $\hat{H}_1=\hat J_{\boldsymbol{n}}$. The near-optimal Dicke superposition state $\lvert D^{(12)}_{5,7}\rangle$, the GHZ state, the
W--superposition state $\lvert D^{(12)}_{1,11}\rangle$, and the balanced Dicke state $\lvert D_{6,6}\rangle$ are compared under
(a) phase damping, (b) amplitude damping, and (c) global depolarization. The HL  and the SNL are indicated for reference.}
\label{ps8}
 \end{figure}

Figures~\ref{fi8} and~\ref{ps8} compare the noise resilience of the near-optimal Dicke-superposition states with that of W--superposition, GHZ, and balanced Dicke states for $N=12$. 
Under phase damping, the Dicke-superposition probe $\lvert D^{(12)}_{5,7}\rangle$ and the balanced Dicke state 
$\lvert D_{6,6}\rangle$ exhibit comparable robustness, maintaining larger QFI over a broad range of noise strengths and outperforming the GHZ and W--superposition probe $\lvert D^{(12)}_{1,11}\rangle$ (see Figs.~2(a), 3(a)). 
Under amplitude damping, the balanced Dicke state maintains phase sensitivity between the SNL and the HL over a broader range of noise strengths than the GHZ, W--superposition, and near-optimal Dicke-superposition states (see Fig.~3(b)). Under global depolarization, the optimal probe state (GHZ) and the near-optimal Dicke-superposition state $\vert D^{(12)}_{5,7}\rangle$ exhibit similar metrological behavior, maintaining phase sensitivity between the HL and SNL across almost the entire range of noise strengths (see Fig.~3(c)).

Under local phase damping, the near-optimal Dicke-superposition probes and the balanced Dicke states exhibit a noticeably slower degradation of metrological performance than the GHZ and W-superposition states, as evidenced by the behaviour of both the QFI and the phase sensitivity shown in Figs.~\ref{fi8}(a) and \ref{ps8}(a). The results suggest that the quantum correlations responsible for metrological enhancement are comparatively more resilient in these probe states under local dephasing. A detailed analytical characterization of the underlying mechanism is beyond the scope of the present work; nevertheless, the numerical results clearly demonstrate the superior resilience of the Dicke-superposition and balanced Dicke probes against phase damping noise.

Unlike local phase and amplitude damping channels, global depolarization acts collectively on the entire multipartite state. Consequently, the noise does not selectively affect particular excitation sectors or coherence structures, leading primarily to an overall reduction of the QFI while largely preserving the qualitative ordering observed in the noiseless case.


\begin{figure}[t]  
 \centering
  \includegraphics[width=0.45\textwidth]{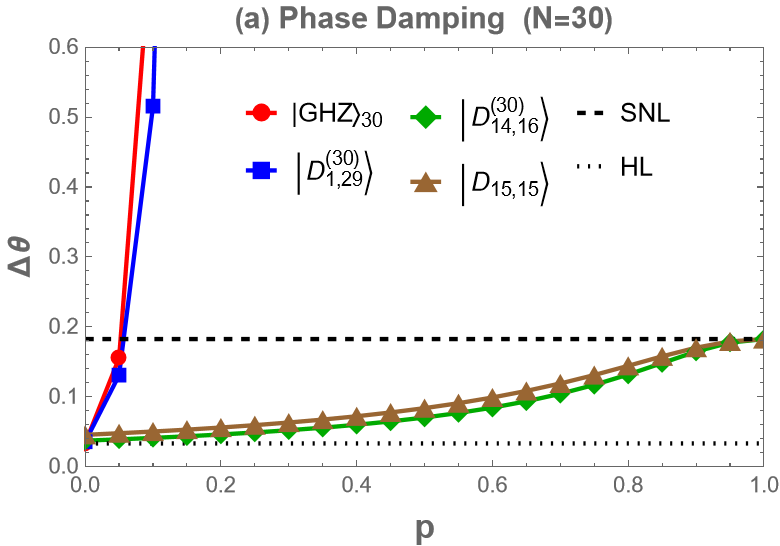} \\[0.3cm]
  \includegraphics[width=0.45\textwidth]{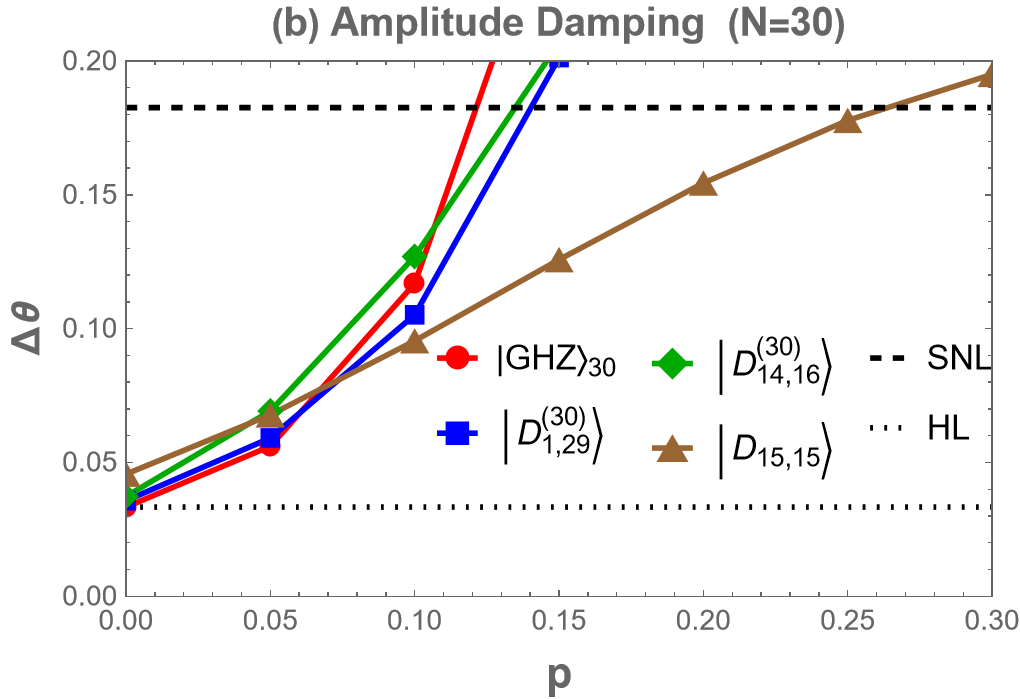} \\[0.3cm]
  \includegraphics[width=0.45\textwidth]{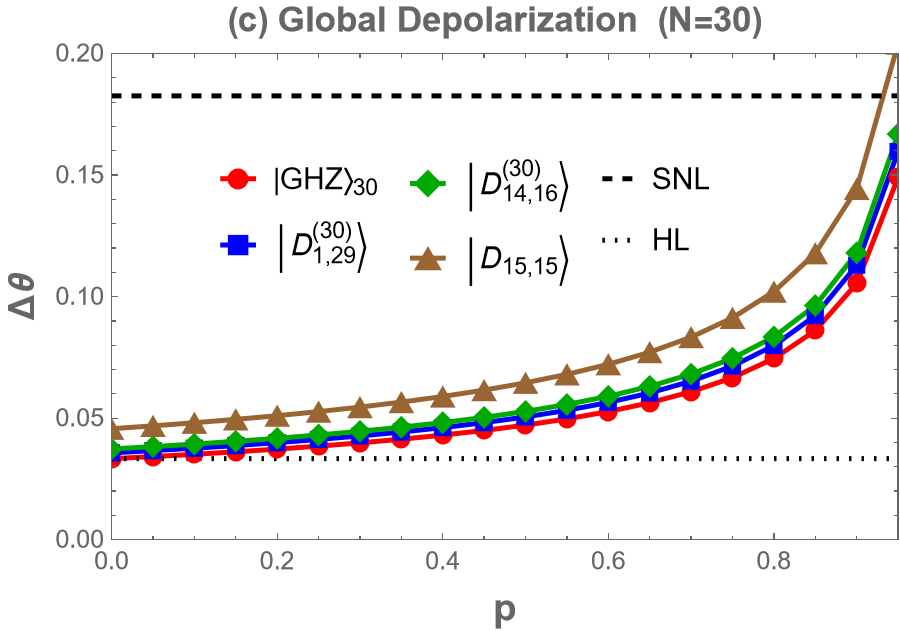} 
  \caption{Phase sensitivity $\Delta\theta$ of  probe states with $N=30$ as a function of the noise parameter $p$ under encoding Hamiltonian $\hat{H}_1=\hat J_{\boldsymbol{n}}$ and subjected to (a) phase damping and (b) amplitude damping 
(c) global depolarization. The HL  and the SNL are indicated for reference.}\label{psnew}
 \end{figure}

As discussed in Sec.~III.A, the near-optimal Dicke superposition states exhibit a larger noiseless QFI than the W-superposition state for system sizes $N\lesssim 20$ (see  Fig.~\ref{1}). However, this ordering is reversed for $N>20$, where the W-superposition state becomes metrologically superior. To examine how this change in the noiseless metrological ordering manifests itself in the presence of noise, Fig.~\ref{psnew} compares the noise resilience of the corresponding probe states for $N=30$ under the three noise models considered.

  Figure~\ref{psnew} shows that although the W-superposition state possesses a higher noiseless QFI than the Dicke superposition state for $N>20$, it continues to remain more vulnerable to phase damping (see Fig.~\ref{psnew}(a)), similar to the behaviour observed for smaller system sizes (see Fig.~\ref{ps8}(a)). In contrast, under global depolarization, the ordering observed in the noiseless case is largely retained, with the W-superposition state exhibiting slightly better phase sensitivity than the Dicke superposition state (see Figs.~\ref{ps8}(c) and \ref{psnew}(c)). These results indicate that the extent to which a noiseless metrological advantage is retained in the presence of noise depends on the nature of the noise channel.

\begin{figure}[ht]
	\centering
		\includegraphics[width=0.45\textwidth]{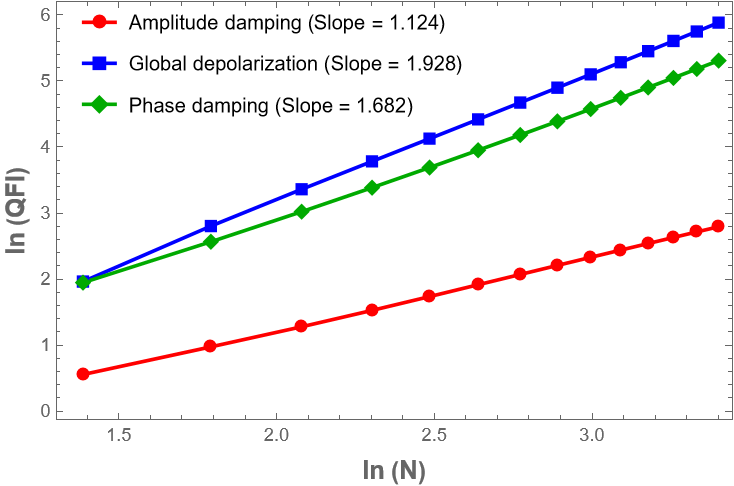}   
		\caption{A log--log plot of the QFI of near-optimal Dicke superposition
states as a function of the number of qubits $N$ at a fixed noise strength
$p=0.5$. Natural logarithms are used for both axes. Results are shown
for phase damping, amplitude damping, and global depolarization. The
slopes characterize the scaling behaviour under the different noise
channels.}	
\label{1a}
\end{figure}

Figure~\ref{1a} displays a log--log plot of the QFI of near-optimal Dicke superposition states as a function of the number of qubits $N$ for the three noisy channels at a fixed noise strength $p=0.5$. Natural logarithms are employed here. The extracted slopes indicate that global depolarization preserves an almost Heisenberg-like scaling, whereas phase damping results in only a moderate reduction of the scaling exponent while retaining substantial quantum enhancement. In contrast, amplitude damping significantly reduces the scaling exponent, driving it closer to the shot-noise limit. The comparatively strong resilience of the near-optimal Dicke superposition states against phase damping is also evident from Fig.~\ref{1a}.

At this juncture, it is worth noting the recent work in
Ref.~\cite{LiRen2026}, where the Dicke superposition states
$\lvert D^{(N)}_{l,l'}\rangle$, with $(l,l')$ specified in
Eqs.~(\ref{odd}) and (\ref{even}), were identified as possessing
near-optimal QFI under encoding by the linear collective-spin Hamiltonian
$\hat{H}_1=\hat J_{\boldsymbol n}$. They investigated the 
noiseless QFI scaling of these states and compared their metrological
performance with that of GHZ and W-superposition states. Thus, there is
a partial overlap between Ref.~\cite{LiRen2026} and the present work.

The scope of the present work, however, differs in several important
respects. First, a systematic comparison of near-optimal Dicke superposition
states with GHZ, W-superposition, and balanced Dicke probes is carried out
in both noiseless and noisy settings. Second, the metrological scaling of
the near-optimal Dicke superposition states under phase damping, amplitude
damping, and global depolarizing noise is analyzed through log--log plots using natural logarithms.
Third, the nonlinear-metrology analysis differs substantially from that of
Ref.~\cite{LiRen2026}. While Ref.~\cite{LiRen2026} considered an interacting
Ising Hamiltonian with nearest-neighbour couplings, the present work
investigates a representative class of four collective two-body interaction
Hamiltonians with all-to-all couplings. The corresponding optimal probe
states are identified, and their metrological scaling under realistic noise
channels is analyzed. In addition, the associated near-optimal Dicke
superposition states are identified and their noise resilience is compared
with that of the corresponding optimal probes.

Having examined the metrological performance and noise resilience of near-optimal Dicke superposition states under linear collective-spin encoding, the following section extends the analysis to parameter encoding generated by collective two-body interaction Hamiltonians.

\section{Metrology with collective two-body encoding Hamiltonians}
Collective two-body interaction Hamiltonians constitute the simplest class of
nonlinear generators in  quantum metrology. Such interactions can generate metrologically useful entanglement during the
encoding process itself, leading to enhanced scaling of the QFI and different classes of optimal probe states~\cite{pezze,hylus,ujwal25,pral25}. The metrological advantages here arise not only from the choice of initial probe state but also from the structure of the encoding Hamiltonian itself. 

The present section focuses on representative  two-body  encoding Hamiltonians of the form $\hat{J}^2_{\boldsymbol{n}}$. Non-linear interactions of this type  underlie the Lipkin--Meshkov--Glick model~\cite{LMG} and one-axis-twisting Hamiltonian~\cite{KitagawaUeda1993}.  
The metrological performance of Dicke superposition probes under such two-body nonlinear encodings is analyzed here. In particular, we identify optimal and near-optimal Dicke superposition probes for these generators and compare their noise resilience.

Consider the following two-body interaction Hamiltonians:
\begin{eqnarray}
\label{h2r}
\hat{H}^{(1)}_2 &=& \eta\,\hat{J}_{\boldsymbol{n}}^2=\frac{\eta}{4}\sum_{i,j=1}^N \hat{\sigma}_{i\boldsymbol{n}}\hat{\sigma}_{j\boldsymbol{n}} \nonumber \\
\hat{H}^{(2)}_2 &=& \mu\,\hat{J}_{\boldsymbol{n}} + \eta\,\hat{J}_{\boldsymbol{n}}^2 \nonumber \\ 
&=&\frac{\mu}{2}\sum_{i=1}^N \hat{\sigma}_{i\boldsymbol{n}}+\frac{\eta}{4}\sum_{i,j=1}^N \hat{\sigma}_{i\boldsymbol{n}}\hat{\sigma}_{j\boldsymbol{n}} \nonumber \\
\hat{H}^{(3)}_2 &=& \frac{\eta}{4}\sum_{i<j}^N \hat{\sigma}_{i\boldsymbol{n}}\hat{\sigma}_{j\boldsymbol{n}}
= \frac{\eta}{2}\left(\hat{J}^2_{\boldsymbol{n}} - \frac{N}{4}\mathbb{I}_{2^N}\right), \nonumber \\
\hat{H}^{(4)}_2 &=& \frac{\mu}{2}\sum_{i=1}^N \hat{\sigma}_{i\boldsymbol{n}} +
\frac{\eta}{4}\sum_{i<j}^N \hat{\sigma}_{i\boldsymbol{n}}\hat{\sigma}_{j\boldsymbol{n}}\nonumber \\
&=& \mu\,\hat{J}_{\boldsymbol{n}}
+\frac{\eta}{2}\left(\hat{J}_{\boldsymbol{n}}^2 - \frac{N}{4}\mathbb{I}_{2^N}\right), 
\end{eqnarray}
where $\mu$ and $\eta$  denote the strengths of the linear and two-body nonlinear interactions, respectively. Throughout this work, we set  $\mu=1=\eta$ for simplicity of analysis. 

The QFI  corresponding to pure optimal input probe states $\vert\psi^{(r)}\rangle_{\rm opt}$ and encoding Hamiltonians  $\hat{H}^{(r)}_2$,\ $r~=~1,2,3,4$ (see Eq.(\ref{h2r})) is given by (see Eq.(\ref{defqfi})) $F_Q\!\left(\lvert\psi^{(r)}\rangle_{\rm opt},\,\hat{H}^{(r)}_2\right)~=~4\, \Delta^2 \hat{H}^{(r)}_2~=~\left(\lambda^{(r)}_{\rm max}-\lambda^{(r)}_{\rm min}\right)^2$, where $\lambda^{(r)}_{\rm max}$ and $\lambda^{(r)}_{\rm min}$ are the respective maximum and minimum eigenvalues of the encoding Hamiltonian $\hat{H}^{(r)}_2$. The optimal probe state is $\vert\psi^{(r)}\rangle_{\rm opt}~=~(|\phi^{(r)}_{\rm max}\rangle ~+~|\phi^{(r)}_{\rm min}\rangle)/\sqrt{2}$ with $|\phi^{(r)}_{\rm max}\rangle,\  |\phi^{(r)}_{\rm min}\rangle$ denoting the eigenstates corresponding respectively to the eigenvalues  $\lambda^{(r)}_{\rm max}$, $\lambda^{(r)}_{\rm min}$ of  $\hat{H}^{(r)}_2$, $r~=~1,2,3,4$.  

Note that the eigenstates of the collective-spin operator $\hat{J}_{\boldsymbol n}$ and its square
$\hat{J}_{\boldsymbol n}^{\,2}$ are the Dicke states $\lvert D_{N-l,l}\rangle_{\boldsymbol n},\ l=0,1,\ldots, N$ with respect to the quantization axis $\hat{\boldsymbol n}$  corresponding respectively to the eigenvalues  $(N/2)-l$, $\left((N/2)-l\right)^2$.  The maximum eigenvalue of $\hat{J}_{\boldsymbol n}$ is $N/2$, while that of
$\hat{J}_{\boldsymbol n}^{\,2}$ is $N^2/4$. The latter is two-fold degenerate,
with the corresponding eigenstates 
\begin{equation} 
\label{01}
\lvert D_{N,0}\rangle_{\boldsymbol n}
\equiv\lvert 0\rangle_{\boldsymbol n}^{\otimes N},\ \  \mbox{and} \ \ \lvert D_{0,N}\rangle_{\boldsymbol n}
\equiv\lvert 1\rangle_{\boldsymbol n}^{\otimes N}.
\end{equation}
Thus, for the encoding Hamiltonian $\hat{H}^{(1)}_2~=~\hat{J}_{\boldsymbol n}^{\,2}$,   
\begin{equation}
\label{ghzn}
\lvert \mathrm{GHZ}\rangle_{\boldsymbol n}
=
\frac{1}{\sqrt{2}}
\left(
\lvert D_{N, 0}\rangle_{\boldsymbol n}
+
\lvert D_{0, N}\rangle_{\boldsymbol n}
\right)
\equiv
\lvert D^{(N)}_{N,0}\rangle_{\boldsymbol n},
\end{equation}
is an eigenstate associated with the maximal eigenvalue $N^2/4$. 

The minimum eigenvalue of $\hat{J}_{\boldsymbol n}$ is $-N/2$, and belongs to the eigenstate
$\lvert D_{0,N}\rangle_{\boldsymbol n}\equiv\lvert 1\rangle_{\boldsymbol n}^{\otimes N}$. In contrast, the minimum eigenvalue of $\hat{J}_{\boldsymbol n}^{\,2}$ depends on
the parity of $N$. For odd $N$, the minimum eigenvalue is $1/4$ with the corresponding eigenstate $\lvert D_{(N-1)/2,(N+1)/2}\rangle_{\boldsymbol n}$.
For even $N$, the minimum eigenvalue is zero, with $\lvert D_{N/2,N/2}\rangle_{\boldsymbol n}$ being the eigenstate.

The eigenvalues $\lambda^{(r)}_{\rm max}$, $\lambda^{(r)}_{\rm min}$, the corresponding eigenvectors $|\phi^{(r)}_{\rm max}\rangle, |\phi^{(r)}_{\rm min}\rangle$ of  $\hat{H}^{(r)}_2$, $r~=~1,2,3,4$ along with  the  QFI $F_Q\!\left(\lvert\psi^{(r)}\rangle_{\rm opt},\,\hat{H}^{(r)}_2\right)$ are summarized in Table~\ref{tab:2}.

The optimal probe states $\lvert\psi^{(r)}\rangle_{\rm opt}$ maximizing the QFI for the 
two-body encoding Hamiltonians $\hat{H}^{(r)}_2,~r~=~1,2,3,4$ (see ~(\ref{h2r})) are given by
\begin{eqnarray}
\label{psiopt}
\lvert\psi^{(1)}_{N={\rm odd}}\rangle&=& \dfrac{1}{\sqrt{2}}\, \left(
\lvert D^{(N)}_{N,0}\rangle_{\boldsymbol n}+\lvert D^{(N)}_{(N-1)/2,(N+1)/2}\rangle_{\boldsymbol n}\right) \nonumber \\ 
\lvert\psi^{(1)}_{N={\rm even}}\rangle&=& \dfrac{1}{\sqrt{2}}
\left(\lvert D^{(N)}_{N,0}\rangle_{\boldsymbol n}+ \lvert D_{N/2,N/2}\rangle_{\boldsymbol n} \right) \nonumber \\
\lvert\psi^{(2)}_{N={\rm odd}}\rangle&=& \dfrac{1}{\sqrt{2}}
\left(\lvert D_{N,0}\rangle_{\boldsymbol n}+\lvert D_{(N-1)/2,(N+1)/2}\rangle_{\boldsymbol n}\right) \nonumber \\ 
\lvert\psi^{(2)}_{N={\rm even}}\rangle&=& \dfrac{1}{\sqrt{2}}
\left(\lvert D_{N,0}\rangle_{\boldsymbol n}+\lvert D^{(N)}_{N/2,(N+2)/2}\rangle_{\boldsymbol{n}}\right)  \\ 
\lvert\psi^{(3)}_{N={\rm odd}}\rangle&=& \lvert\psi^{(1)}_{N={\rm odd}}\rangle \nonumber \\ 
\lvert\psi^{(3)}_{N={\rm even}}\rangle&=& \lvert\psi^{(1)}_{N={\rm even}}\rangle  \nonumber \\ 
\lvert\psi^{(4)}_{N={\rm odd}}\rangle&=& \dfrac{1}{\sqrt{2}}
\left(\lvert D_{N,0}\rangle_{\boldsymbol n} + \lvert D^{(N)}_{(N+1)/2,(N+3)/2}\rangle_{\boldsymbol{n}}\right) \nonumber \\ 
\lvert\psi^{(4)}_{N={\rm even}}\rangle&=& \dfrac{1}{\sqrt{2}}
\left(\lvert D_{N,0}\rangle_{\boldsymbol n}+ \lvert D_{(N-2)/2,(N+2)/2}\rangle_{\boldsymbol{n}} \right) \nonumber 
\end{eqnarray}
\begin{table*}[t]
\begin{center}
\caption{Eigenvalues $\lambda^{(r)}_{\rm max}$, $\lambda^{(r)}_{\rm min}$, corresponding eigenstates
$\lvert\phi^{(r)}_{\rm max}\rangle$, $\lvert\phi^{(r)}_{\rm min}\rangle$ of the two-body encoding Hamiltonians 
$\hat{H}^{(r)}_2$ $(r=1,2,3,4)$, along with the resulting quantum Fisher information 
$F_Q\!\left(\lvert\psi^{(r)}\rangle_{\rm opt},\,\hat{H}^{(r)}_2\right)$. 
Throughout the table, Dicke states $\lvert D_{N-l,l}\rangle_{\boldsymbol{n}}$, Dicke superposition states $\lvert D^{(N)}_{l,l'}\rangle_{\boldsymbol{n}}$
are defined with  respect to the quantization axis
 $\hat{\boldsymbol{n}}$.}
\medskip
\footnotesize
\renewcommand{\arraystretch}{1.6}
\setlength{\tabcolsep}{6pt}
\begin{tabular}{c c c c c}
\toprule
 & $\hat H^{(1)}_2=\hat{J}_{\boldsymbol{n}}^2$ 
 & $\hat H^{(2)}_2=\hat{J}_{\boldsymbol{n}} + \hat{J}_{\boldsymbol{n}}^2$
 & $\hat H^{(3)}_2= \frac{1}{2}\left(\hat{J}^2_{\boldsymbol{n}} - \frac{N}{4}\,\mathbb{I}_{2^N}\right)$
 & $\hat H^{(4)}_2= \hat{J}_{\boldsymbol{n}}+\frac{1}{2}\left(\hat{J}_{\boldsymbol{n}}^2 - \frac{N}{4}\,\mathbb{I}_{2^N}\right)$ \\
\midrule
$\lambda^{(r)}_{\rm max}$ 
& $N^2/4$ 
& $N(N+2)/4$ 
& $N(N-1)/8$ 
& $N(N+3)/8$ \\[6pt]
\midrule
$\lambda^{(r)}_{\rm min}$ (odd $N$) 
& $1/4$ 
& $-1/4$ 
& $-(N-1)/8$ 
& $-(N+3)/8$ \\[6pt]
\midrule
$\lambda^{(r)}_{\rm min}$ (even $N$) 
& $0$ 
& $0$ 
& $-N/8$ 
& $-(N+4)/8$ \\[6pt]
\midrule
$\lvert\phi^{(r)}_{\rm max}\rangle$ 
&  $\lvert D^{(N)}_{N,0}\rangle_{\boldsymbol{n}}$ 
&  $\lvert D_{N,0}\rangle_{\boldsymbol{n}}$ 
&  $\lvert D^{(N)}_{N,0}\rangle_{\boldsymbol{n}}$
&  $\lvert D_{N,0}\rangle_{\boldsymbol{n}}$ \\[6pt]
\midrule
$\lvert\phi^{(r)}_{\rm min}\rangle$ (odd $N$) 
& $\lvert D^{(N)}_{(N-1)/2,(N+1)/2}\rangle_{\boldsymbol{n}}$ 
& $\lvert D_{(N-1)/2,(N+1)/2}\rangle_{\boldsymbol{n}}$ 
& $\lvert D^{(N)}_{(N-1)/2,(N+1)/2}\rangle_{\boldsymbol{n}}$
& $\lvert D^{(N)}_{(N+1)/2,(N+3)/2}\rangle_{\boldsymbol{n}}$ \\[6pt]
\midrule
$\lvert\phi^{(r)}_{\rm min}\rangle$ (even $N$) 
& $\lvert D_{N/2,N/2}\rangle_{\boldsymbol{n}}$ 
& $\lvert D^{(N)}_{N/2, (N+2)/2}\rangle_{\boldsymbol{n}}$ 
& $\lvert D_{N/2,N/2}\rangle_{\boldsymbol{n}}$ 
& $\lvert D_{(N-2)/2,(N+2)/2}\rangle_{\boldsymbol{n}}$ \\[6pt]
\midrule
$F_Q$ (odd $N$) 
& $(N^2-1)^2/16$ 
& $(N+1)^4/16$ 
& $(N^2-1)^2/64$ 
& $(N+3)^2(N+1)^2/64$ \\[6pt]
\midrule
$F_Q$ (even $N$) 
& $N^4/16$ 
& $N^2(N+2)^2/16$ 
& $N^4/64$ 
& $(N+2)^4/64$ \\
\bottomrule
\end{tabular}
\label{tab:2}
\end{center}
\end{table*}
For metrological schemes based on nonlinear $k$-body generators, the precision
benchmarks differ markedly from those of linear encodings~\cite{prl07,shaji08, Boixo2008}. While optimal separable probes
yield~\cite{pral25} a Fisher information scaling ${\cal O}(N^{2k-1})$,  entangled states
can achieve the  scaling~\cite{prl07,shaji08, Boixo2008,pral25,ujwal25}\  ${\cal O}(N^{2k})$, corresponding to a
separable state precision bound $\Delta\theta \sim N^{-(k-\tfrac{1}{2})}$ and the
nonlinear Heisenberg limit $\Delta\theta \sim N^{-k}$.  For the present case of collective two-body encoding Hamiltonians ($k=2$), these general scalings reduce to
$F_Q^{\rm sep} = \mathcal{O}(N^3)$ and $F_Q^{\rm ent} = \mathcal{O}(N^4)$,
corresponding respectively to the nonlinear shot-noise limit
$\Delta\theta \sim N^{-3/2}$ and the nonlinear Heisenberg limit $\Delta\theta \sim N^{-2}$.

These two regimes are denoted here as the
nonlinear shot-noise limit (NL-SNL) and the nonlinear Heisenberg limit 
(NL-HL), respectively. A probe state
whose phase sensitivity satisfies
\begin{equation}
(\Delta \theta)_{\rm NL\text{-}HL} \leq \Delta \theta < (\Delta \theta)_{\rm NL\text{-}SNL}
\end{equation}
exhibits a genuine metrological advantage over separable probes and
constitutes a practically useful quantum resource.

Table~\ref{tab:NLsnl-hl} summarizes the nonlinear metrological performance of the two-body interaction Hamiltonians $\hat{H}^{(r)}_2$, $r=1,2,3,4$ for $N=12$.   The nonlinear shot-noise limit is determined by the maximal Fisher information
$F_Q\!\left(|\psi_{\rm sep}^{\rm opt}\rangle,\hat{H}^{(r)}_2\right)$ of the optimal product state
$|\psi_{\rm sep}^{\rm opt}\rangle
=\left(\cos\theta_{\rm opt}|0\rangle+\sin\theta_{\rm opt}|1\rangle\right)^{\otimes N}$. The values of $F_Q\!\left(|\psi_{\rm sep}^{\rm opt}\rangle,\hat{H}^{(r)}_2\right)$ obtained after numerical optimization are listed in the second column of Table~\ref{tab:NLsnl-hl} and they specify 
the NL-SNL values for each Hamiltonian $\hat{H}^{(r)}_2$, $r=1,2,3,4$, for the system size $N=12$.

\begin{table}[t]
\centering
\caption{Nonlinear metrological benchmarks for two-body Hamiltonians $\hat{H}^{(r)}_2$, $r=1,2,3,4$ (see Eqs.~(\ref{h2r})) for $N=12$. The QFI $F_Q(|\psi\rangle_{\mathrm{sep}}, \hat{H}^{(r)}_2)$ achievable by separable (product) states, the QFI $F_Q(|\psi^{(r)}_{(N=12)}\rangle_{\boldsymbol{n}}, \hat{H}^{(r)}_2)$ for the optimal Dicke superposition probe $|\psi^{(r)}_{(N=12)}\rangle_{\boldsymbol{n}}$, and the corresponding phase sensitivities $(\Delta \theta)_{\rm NL\text{-}SNL}$, $(\Delta \theta)_{\rm NL\text{-}HL}$ obtained via the standard Cramér–Rao bound $\Delta\theta = 1/\sqrt{F_Q}$ are reported.}
\scriptsize
\begin{ruledtabular}
\begin{tabular}{ccccc}
$\hat{H}^{(r)}_2$ & $F_Q(|\psi\rangle_{\mathrm{sep}}, \hat{H}^{(r)}_2)$ & $F_Q(|\psi^{(r)}_{(N=12)}\rangle_{\boldsymbol{n}}, \hat{H}^{(r)}_2)$ & $(\Delta \theta)_{\rm NL\text{-}SNL}$ & $(\Delta \theta)_{\rm NL\text{-}HL}$ \\
\hline
$\hat{H}^{(1)}_2$ & 380.29  & 1296 & 0.051 & 0.028   \\
& & & & \\
$\hat{H}^{(2)}_2$ & 483.35  & 1764 & 0.045 & 0.024 \\
& & & & \\
$\hat{H}^{(3)}_2$ & 95.07  & 324 & 0.102 & 0.055 \\
& & & & \\
$\hat{H}^{(4)}_2$ & 150.55  & 600.25 & 0.081 & 0.041 
\end{tabular}
\end{ruledtabular}
\label{tab:NLsnl-hl}
\end{table}

\subsection{Near-optimal Dicke superposition probes for nonlinear two-body encodings}

Having identified the specific class of Dicke superposition states that exhibit near-optimal phase sensitivity in linear (non-interacting) metrology, it is natural to explore the nature of near-optimal states under two-body interaction encodings and to analyze whether they remain metrologically advantageous in noisy environments.

 For encoding dynamics generated by the collective two-body Hamiltonians
$\hat H^{(r)}_2$ ($r=1,2,3,4$) (see Eqs.~(\ref{h2r})),
the Fisher information associated with each Dicke superposition state
of a fixed system size $N$ is optimized over the direction
$\boldsymbol n$, thereby yielding its maximal Fisher information and
the corresponding optimal direction. The Dicke superposition state
whose maximal Fisher information is closest to the optimal QFI value
is designated as the near-optimal probe for the given system size $N$
and Hamiltonian $\hat H^{(r)}_2$ ($r=1,2,3,4$).
The maximal Fisher information values and the corresponding near-optimal Dicke superposition probe states for system sizes $N=10$--$14$ are listed in Table~\ref{tab:near-optimal}. Table~\ref{tab:nearoptimal2} summarizes the general forms of the near-optimal Dicke superposition state families associated with the two-body interaction Hamiltonians 
$\hat{H}^{(r)}_2$ ($r=1,2,3,4$).
A comparison of Tables~\ref{tab:near-optimal} and~\ref{tab:nearoptimal2} with Table~\ref{tab:dicke_superposition_qfi} shows that the near-optimal states for two-body interaction metrology differ qualitatively from those identified in the linear metrology regime (see Eqs.~(\ref{sdo}) and~(\ref{sde})).
\begin{table*}[t]
\begin{center}
\caption{
The QFI of near-optimal Dicke superposition probe states $\lvert D^{(N)}_{l,l'}\rangle_{\boldsymbol{n}} \equiv
(\lvert D_{N-l,l}\rangle_{\boldsymbol{n}} + \lvert D_{N-l',l'}\rangle)_{\boldsymbol{n}}/\sqrt{2}$, for  values of $N=10$ to $N=14$, with respect to quantization axis $\hat{\boldsymbol{n}}$
for  collective two-body encoding Hamiltonians $\hat{H}^{(r)}_2$ $(r=1,\,2,\,3,\,4)$.}
\medskip
\footnotesize
\renewcommand{\arraystretch}{1.6}
\setlength{\tabcolsep}{6pt}
\begin{tabular}{c c c c c}
\toprule
 & $\hat{H}^{(1)}_2=\hat{J}_{\boldsymbol{n}}^2$ & $\hat{H}^{(2)}_2=\hat{J}_{\boldsymbol{n}} + \hat{J}_{\boldsymbol{n}}^2$ & $\hat{H}^{(3)}_2= \frac{1}{2}\left(\hat{J}^2_{\boldsymbol{n}} - \frac{N}{4}\,\mathbb{I}_{2^N}\right)$ & $\hat{H}^{(4)}_2=\hat{J}_{\boldsymbol{n}}+\frac{1}{2}\left(\hat{J}_{\boldsymbol{n}}^2 - \frac{N}{4}\mathbb{I}_{2^N}\right)$ \\
\midrule
\multirow{1}{*} {$N=10$}
 & $F_Q(\lvert D^{(10)}_{0,6}\rangle_{\boldsymbol{n}}, \hat{H}^{(1)}_2)\simeq 576$ & $F_Q(\lvert D^{(10)}_{1,3}\rangle_{\boldsymbol{n}}, \hat{H}^{(2)}_2)\simeq 819$ & $F_Q(\lvert D^{(10)}_{0,6}\rangle_{\boldsymbol{n}}, \hat{H}^{(3)}_2)\simeq 144$
 & $F_Q(\lvert D^{(10)}_{0,5}\rangle_{\boldsymbol{n}}, \hat{H}^{(4)}_2)\simeq 306$ \\
\midrule
\multirow{1}{*}{$N=11$}
& $F_Q(\lvert D^{(11)}_{1,3}\rangle_{\boldsymbol{n}}, \hat{H}^{(1)}_2)\simeq 809$ & $F_Q(\lvert D^{(11)}_{0,5}\rangle_{\boldsymbol{n}}, \hat{H}^{(2)}_2)\simeq 1225$ & $F_Q(\lvert D^{(11)}_{1,3}\rangle_{\boldsymbol{n}}, \hat{H}^{(3)}_2)\simeq 202$
 & $F_Q(\lvert D^{(11)}_{1,3}\rangle_{\boldsymbol{n}}, \hat{H}^{(4)}_2)\simeq 408$     \\
\midrule
\multirow{1}{*}{$N=12$}
& $F_Q(\lvert D^{(12)}_{0,5}\rangle_{\boldsymbol{n}}, \hat{H}^{(1)}_2)\simeq 1225$ & $F_Q(\lvert D^{(12)}_{1,3}\rangle_{\boldsymbol{n}}, \hat{H}^{(2)}_2)\simeq 1625$ & $F_Q(\lvert D^{(12)}_{0,5}\rangle_{\boldsymbol{n}}, \hat{H}^{(3)}_2)\simeq 306$
 & $F_Q(\lvert D^{(12)}_{0,6}\rangle_{\boldsymbol{n}}, \hat{H}^{(4)}_2)\simeq 576$     \\
\midrule
\multirow{1}{*}{$N=13$}
& $F_Q(\lvert D^{(13)}_{1,3}\rangle_{\boldsymbol{n}}, \hat{H}^{(1)}_2)\simeq 1617$ & $F_Q(\lvert D^{(13)}_{0,6}\rangle_{\boldsymbol{n}}, \hat{H}^{(2)}_2)\simeq 2304$ & $F_Q(\lvert D^{(13)}_{1,3}\rangle_{\boldsymbol{n}}, \hat{H}^{(3)}_2)\simeq 404$
 & $F_Q(\lvert D^{(13)}_{0,6}\rangle_{\boldsymbol{n}}, \hat{H}^{(4)}_2)\simeq 729$     \\ 
 \midrule
\multirow{1}{*}{$N=14$}
& $F_Q(\lvert D^{(14)}_{0,6}\rangle_{\boldsymbol{n}}, \hat{H}^{(1)}_2)\simeq 2304$ & $F_Q(\lvert D^{(14)}_{0,6}\rangle_{\boldsymbol{n}}, \hat{H}^{(2)}_2)\simeq 2916$ & $F_Q(\lvert D^{(14)}_{0,6}\rangle_{\boldsymbol{n}}, \hat{H}^{(3)}_2)\simeq 575$
 & $F_Q(\lvert D^{(14)}_{0,7}\rangle_{\boldsymbol{n}}, \hat{H}^{(4)}_2)\simeq 992$     \\
 \bottomrule
\end{tabular}
\label{tab:near-optimal}
\end{center}
\end{table*} 

\begin{table*}[t]
\caption{Near-optimal Dicke superposition states corresponding to two-body interaction Hamiltonians $\hat H^{(r)}_2, r=1,\,2,\,3,\,4$.}
\label{tab:nearoptimal2}
\begin{ruledtabular}
\begin{tabular}{c|c|c|c|c}
 & 
$\hat{H}^{(1)}_2=\hat{J}_{\boldsymbol n}^{\,2}$ 
&
$\hat{H}^{(2)}_2=\hat{J}_{\boldsymbol n}+\hat{J}_{\boldsymbol n}^{\,2}$ 
&
$\hat{H}^{(3)}_2=\dfrac{1}{2}\left(\hat{J}_{\boldsymbol n}^{\,2}-\dfrac{N}{4}\,\mathbb{I}_{2^N}\right)$
&
$\hat{H}^{(4)}_2=\hat{J}_{\boldsymbol n}
+\dfrac{1}{2}\left(\hat{J}_{\boldsymbol n}^{\,2}-\dfrac{N}{4}\,\mathbb{I}_{2^N}\right)$
\\[8pt]
\hline
Odd $N$
&
$\displaystyle 
\left|D^{(N)}_{0,\frac{N\pm3}{2}}\right\rangle,
\quad N\geq15$
&
$\displaystyle 
\left|D^{(N)}_{0,\frac{N+1}{2}\pm1}\right\rangle,
\quad N\geq9$
&
$\displaystyle 
\left|D^{(N)}_{0,\frac{N\pm3}{2}}\right\rangle,
\quad N\geq15$
&
$\displaystyle 
\left|D^{(N)}_{0,\frac{N\pm3}{2}}\right\rangle,
\quad N\geq13$
\\[12pt]

\cline{1-5}

Even $N$
&
$\displaystyle 
\left|D^{(N)}_{0,\frac{N}{2}\pm 1}\right\rangle,
\quad N\geq10$
&
$\displaystyle 
\left|D^{(N)}_{0,\frac{N-2}{2}}\right\rangle,\
\left|D^{(N)}_{0,\frac{N+4}{2}}\right\rangle,
\quad N\geq14$
&
$\displaystyle 
\left|D^{(N)}_{0,\frac{N}{2}\pm 1}\right\rangle,
\quad N\geq10$
&
$\displaystyle 
\left|D^{(N)}_{0,\frac{N}{2}\pm1}\right\rangle,
\quad N\geq8$
\end{tabular}
\end{ruledtabular}
\end{table*}
Noise robustness of the optimal probe states and the near-optimal Dicke superposition states (encoded by the nonlinear two-body interaction Hamiltonians $\hat{H}^{(r)}_2$,~$r~=~1,2,3,4$)  is investigated in the following subsection.

\subsection{Noise resilience of optimal and near-optimal Dicke superposition probes in two-body metrology}
The noise robustness of the optimal and near-optimal Dicke superposition probe states via two-body interaction Hamiltonians is now analyzed under
realistic decoherence mechanisms. In particular,  metrological performance  of these probes under local phase damping, local amplitude damping, and global depolarizing noise is studied. 
\begin{figure}[t]
	\centering
		 \includegraphics[width=0.45\textwidth]{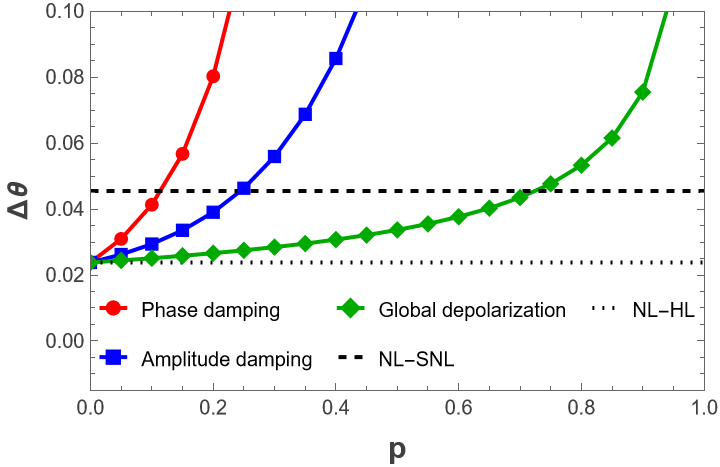}
		\caption{Comparison of phase sensitivity $\Delta \theta$ of $N=12$ optimal probe  $\lvert \psi_{(N=12)}^{(2)}\rangle_{\boldsymbol{n}}=\dfrac{1}{\sqrt{2}}
\left(\lvert D_{12,0}\rangle_{\boldsymbol n}+\lvert D^{(12)}_{6,7}\rangle_{\boldsymbol n}\right)$ (see Eq.(\ref{psiopt})) with parameter encoding generated by the two-body interaction Hamiltonian $\hat{H}^{(2)}_2=\hat{J}_{\boldsymbol{n}} + \hat{J}_{\boldsymbol{n}}^2$.  The plots indicate the relative robustness of the optimal probe  under  phase damping, amplitude damping, and global depolarization. The NL-SNL and the NL-HL are also indicated.}
\label{optallnl}
\end{figure} 
\begin{figure}[ht]
	\centering
		\includegraphics[width=0.45\textwidth]{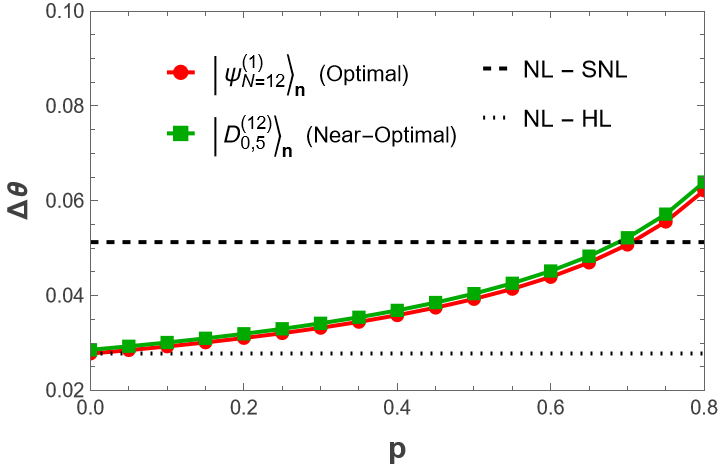}
		 \caption{Comparison of phase sensitivity  $\Delta \theta$ for the  optimal probe $\lvert \psi_{(N=12)}^{(1)}\rangle_{\boldsymbol{n}}=\dfrac{1}{\sqrt{2}}
\left(\lvert D^{(12)}_{12,0}\rangle_{\boldsymbol n}+\lvert D_{6,6}\rangle_{\boldsymbol n}\right)$ with that of near-optimal Dicke superposition probe  $\vert D^{(12)}_{0,5}\rangle_{\boldsymbol{n}}$, both encoded via the two-body interaction Hamiltonian
		$\hat{H}^{(1)}_{2}=\hat{J}^2_{\boldsymbol{n}}$,  subjected to global depolarization. The NL-SNL and the NL-HL are displayed for reference. The optimal and near-optimal states corresponding to the other three two-body interaction Hamiltonians $\hat{H}^{(r)}_2$, $r=2,3,4$ exhibit analogous response to global depolarizing noise.} 
        \label{gdjnl}
		\end{figure} 
The impact of these  three noise processes on the Fisher information of the identified probe states is summarized in Figs.~\ref{optallnl}--\ref{ampnl}. Figure~\ref{optallnl} displays the phase sensitivity of the optimal  probe state $\lvert \psi_{(N=12)}^{(2)}\rangle_{\boldsymbol{n}}~=~\dfrac{1}{\sqrt{2}}
\left(\lvert D_{12,0}\rangle_{\boldsymbol n}+\lvert D^{(12)}_{6,7}\rangle_{\boldsymbol n}\right)$ for $N=12$, subjected to  three different noises, under encoding by the  
  Hamiltonian $\hat{H}^{(2)}_2~=~\hat{J}_{\boldsymbol n} + \hat{J}_{\boldsymbol n}^2$. The dependence of the phase sensitivity on the noise parameter 
$p$ follows a similar trend for the remaining three Hamiltonians $\hat{H}^{(r)}_2$, $r=1,\,3,\,4$.   It is observed, in general, that the optimal probes 
 are particularly susceptible to local phase damping and amplitude damping while exhibiting comparatively greater resilience against 
global depolarization. 

Figs.~\ref{gdjnl}--\ref{ampnl} present a direct comparison between the optimal probe states (see Eq.~(\ref{psiopt}), evaluated for the corresponding encoding and noise model) and the near-optimal Dicke superposition probes listed in Table~\ref{tab:near-optimal}. These results highlight noise regimes in which the near-optimal Dicke superposition probes display metrological performance comparable to, and in certain cases outperforming that of the corresponding optimal states under decoherence. 

 It follows from Figs.~\ref{pdnl}, \ref{ampnl} that, unlike the linear collective-spin encoding scenario, the optimal and near-optimal  probes corresponding to two-body interaction Hamiltonians retain comparable metrological performance only in the weak-noise regime ($p \lesssim 0.3$) under local noisy channels. This behavior is qualitatively consistent with the general observation that nonlinear encoding schemes are particularly susceptible to noise-induced error amplification~\cite{Lordi2025NoiseConstraints}. In contrast, under global depolarization, which acts collectively on the multipartite state, the performance of both probe classes remains comparable (see Fig.~\ref{gdjnl}).

\begin{figure*}[htb]
	\centering
		\includegraphics[width=0.45\textwidth]{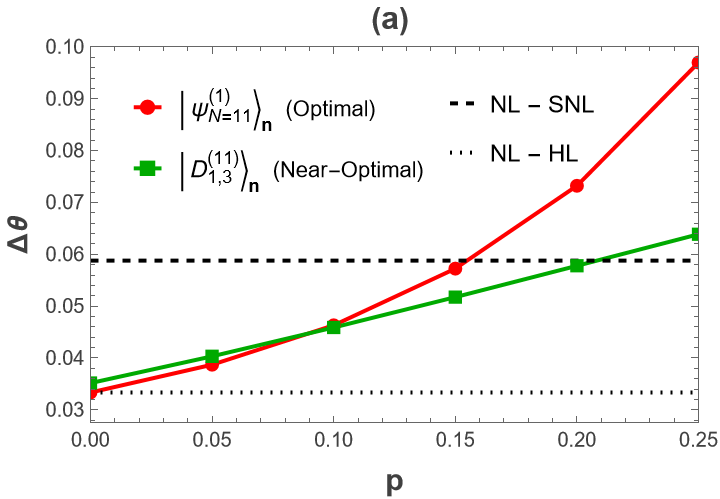}
		\includegraphics[width=0.45\textwidth]{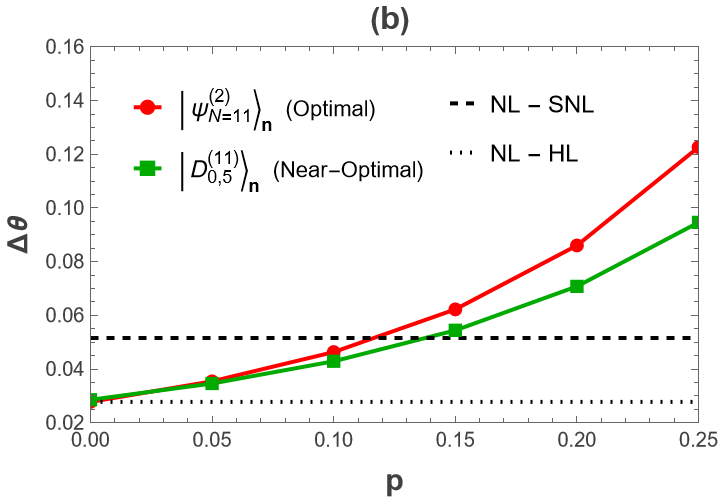}\\[0.3cm]
		\includegraphics[width=0.45\textwidth]{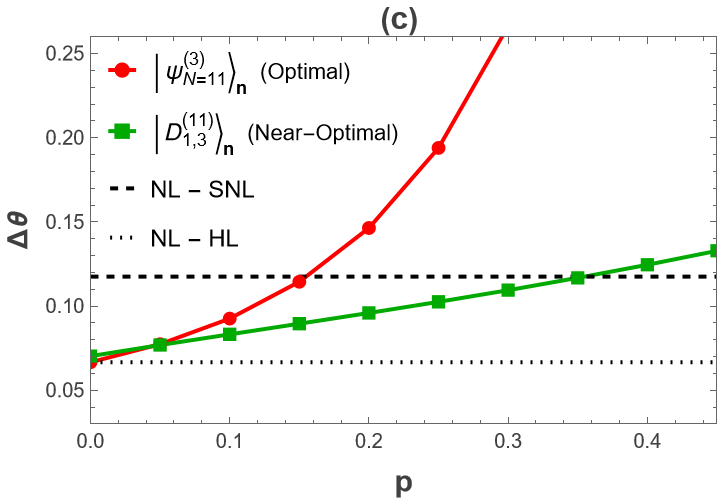}
		\includegraphics[width=0.45\textwidth]{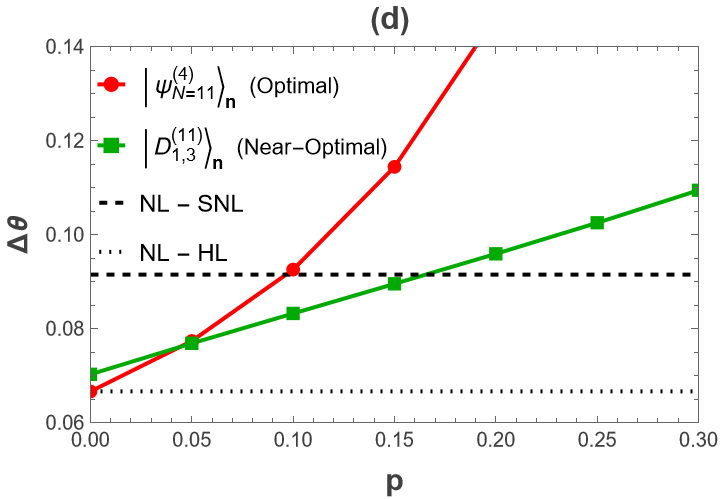}
		 \caption{Comparison of phase sensitivity of  optimal probes $\lvert \psi_{(N=11)}^{(r)}\rangle_{\boldsymbol{n}}$ with that of near-optimal Dicke superposition probes (see Table~\ref{tab:near-optimal}) corresponding to the two-body interaction Hamiltonians  $\hat{H}^{(r)}_2$, $r=1,\,2,\,3,\,4$, under phase damping. References in terms of NL-SNL and the NL-HL are indicated.} \label{pdnl} 
		\end{figure*} 
\begin{figure*}[htb]
	\centering
		\includegraphics[width=0.45\textwidth]{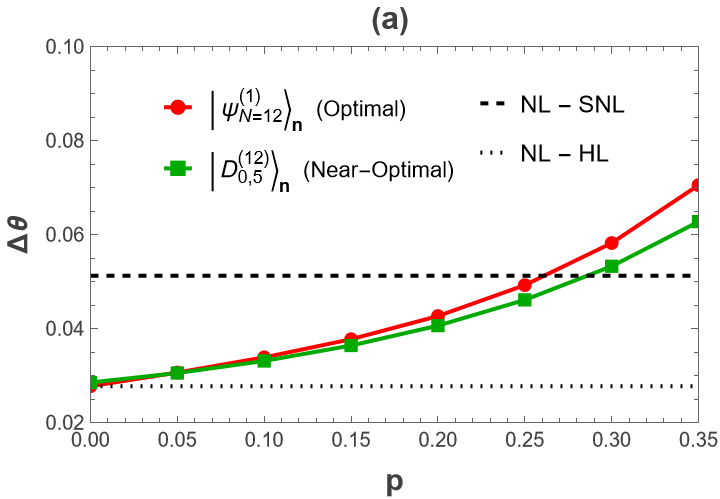}
		\includegraphics[width=0.45\textwidth]{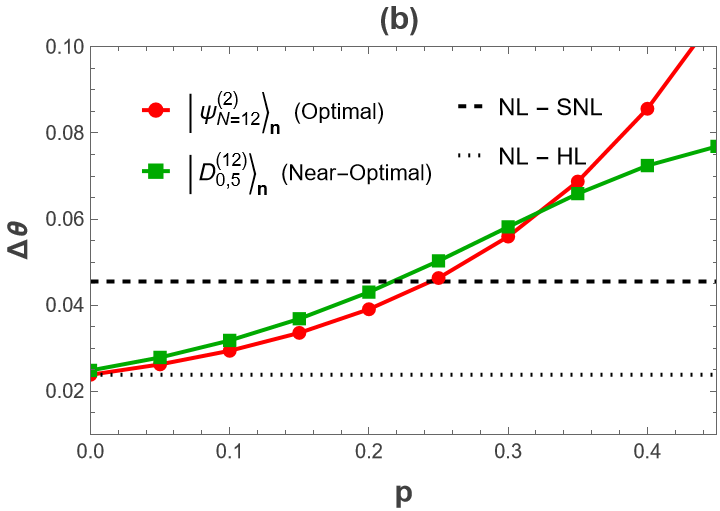}\\[0.3cm]
		\includegraphics[width=0.45\textwidth]{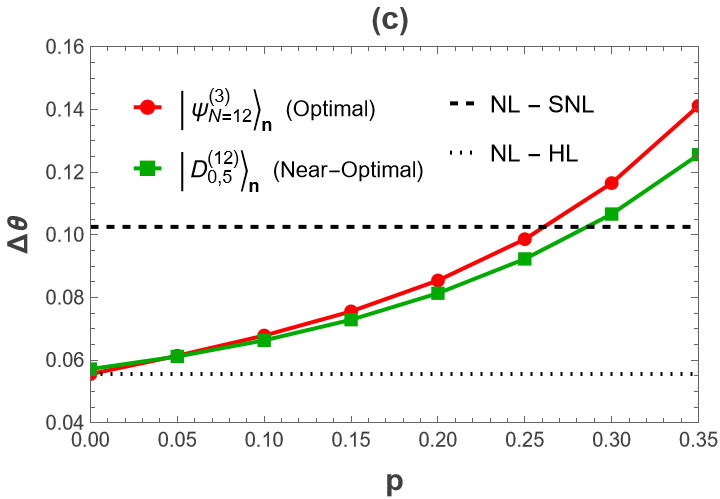}
		\includegraphics[width=0.45\textwidth]{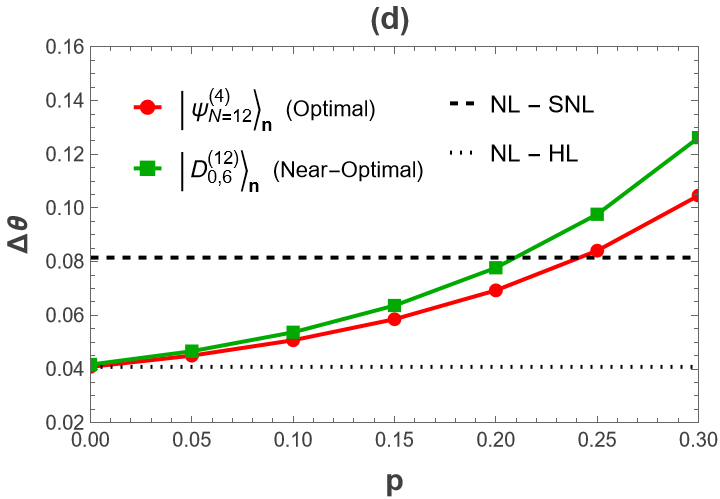}
		 \caption{Comparison of phase sensitivity of $N=12$ optimal probes $\lvert \psi_{(N=12)}^{(r)}\rangle_{\boldsymbol{n}}$ with that of near-optimal Dicke superposition probes (see Table~\ref{tab:near-optimal}), corresponding to the two-body interaction Hamiltonians  $\hat{H}^{(r)}_2$, $r=1,\,2,\,3,\,4$, under amplitude damping. \label{ampnl}}
		\end{figure*}
\begin{figure}[ht]
	\centering
		\includegraphics[width=0.45\textwidth]{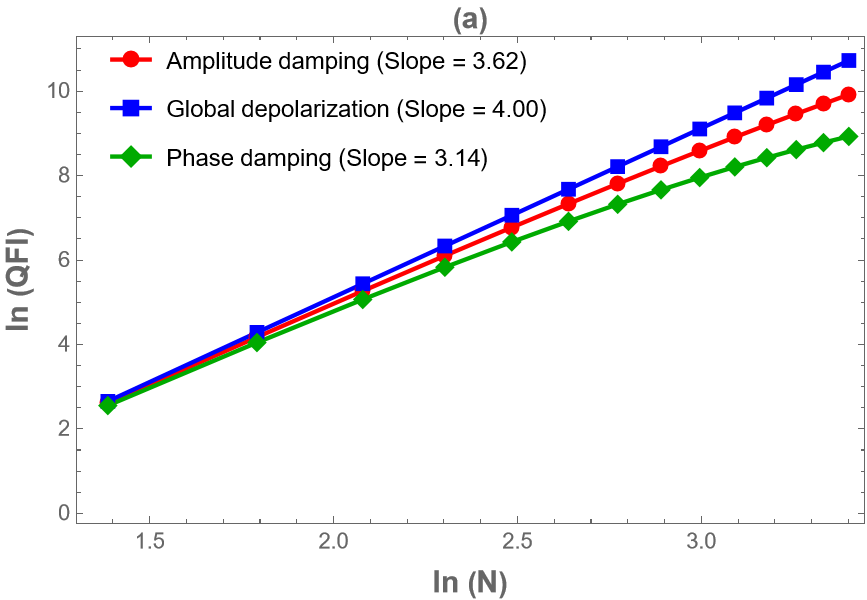} \\[0.2cm]
        \includegraphics[width=0.45\textwidth]{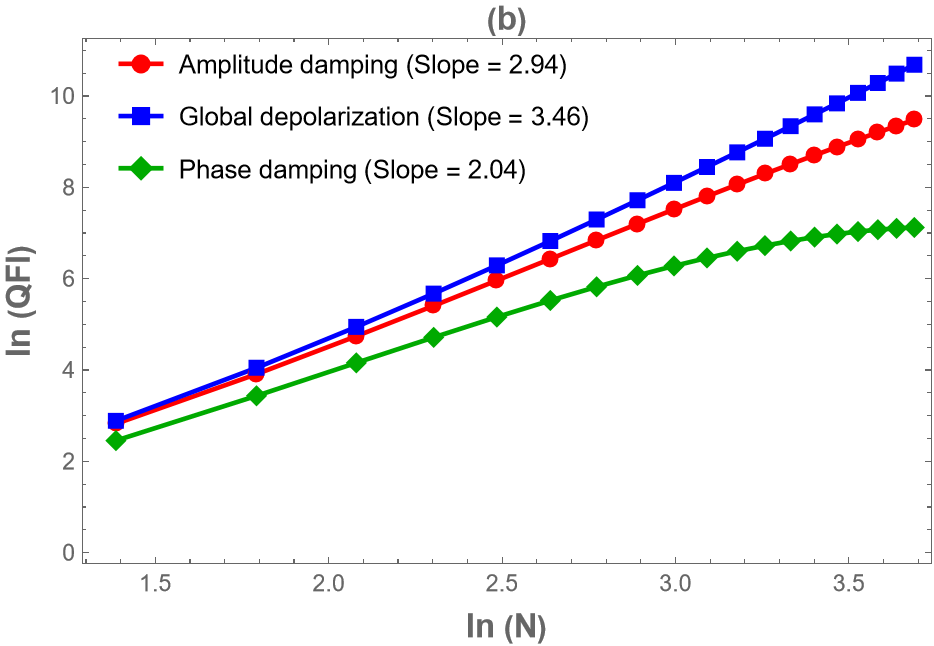}   
		\caption{A log--log variation  of the QFI with the number of qubits $N$ for the optimal probe states encoded by the nonlinear collective interaction Hamiltonians (a) $\hat{H}^{(1)}_2=\hat{J}^2_{\boldsymbol n}$ and (b) 
        $\hat{H}^{(4)}_2=\hat{J}_{\boldsymbol n}
+\dfrac{1}{2}\left(\hat{J}_{\boldsymbol n}^{\,2}-\dfrac{N}{4}\,\mathbb{I}_{2^N}\right)$ in the presence of global depolarization, phase damping, and amplitude damping noise channels for a fixed noise parameter $p=0.1$. Natural logarithms are used for both axes. The scaling exponents obtained from linear fits are shown in the legend.}	
\label{llh1h4}
\end{figure}

 
In order to further characterize the metrological behavior of the optimal Dicke superposition probes  
in the  two-body interaction encoding,  a scaling analysis 
of their QFI is carried out at a fixed noise strength $p=0.1$, corresponding to the weak-noise regime. 
A log--log variation of the QFI of the optimal probe states corresponding to the nonlinear interaction Hamiltonians $\hat{H}_2^{(1)}$ and $\hat{H}_2^{(4)}$ under the three noisy channels is shown in Fig.~\ref{llh1h4}, where natural logarithms are used for both the axes.  The extracted  slopes characterize the finite-size scaling behavior of the QFI. In particular, global depolarization largely preserves an approximately quartic scaling, with the slope remaining close to $4$, while amplitude damping reduces the scaling exponent, although a substantial metrological advantage is still retained for the Hamiltonians considered (see Fig.~\ref{llh1h4}).
Phase damping leads to the strongest suppression among the three noise channels; nevertheless, the scaling remains above the Heisenberg limit within the examined range. These results indicate that, under weak noise, the optimal states associated with the nonlinear Hamiltonians retain quantum-enhanced metrological scaling beyond the standard Heisenberg limit.

Overall, these results demonstrate that, even in the presence of realistic noise,
carefully chosen Dicke superposition probes can retain a  metrological
advantage over standard benchmarks, while offering a favorable trade-off between
phase sensitivity and robustness in nonlinear two-body interaction metrology. 
\section{Conclusions}
This work has investigated the metrological performance of Dicke superposition
probe states in quantum phase estimation protocols governed by linear and
nonlinear (two-body) collective-spin Hamiltonians, with a particular emphasis on
noise resilience under realistic decoherence mechanisms.
Building on the near-optimality of noise-resilient Dicke superposition probes in linear
(non-interacting) metrology, a systematic extension to collective two-body
interaction generators is carried out. For four representative two-body encoding Hamiltonians
$\hat{H}^{(r)}_{2}$, $r=1,2,3,4$, the maximum and minimum eigenvalues,
the corresponding eigenstates, and the resulting QFI are determined
analytically, allowing for the identification of optimal probe states in the
noiseless setting. The associated scaling behavior confirms the emergence of
nonlinear metrological limits, with separable probes bounded by a nonlinear
SNL scaling and entangled probes capable of attaining the nonlinear
HL scaling characteristic of two-body interactions.  Near-optimal Dicke superposition probe states were identified  
by explicitly evaluating and optimizing the Fisher information.
The robustness of both optimal and near-optimal probes has then been examined under local phase damping, local amplitude damping, and global depolarizing noise.  Their relative metrological performance under these decoherence channels has been analyzed for the system sizes under consideration.

In summary, appropriately chosen Dicke superposition states constitute useful metrological resources in both linear and nonlinear (two-body interaction) quantum metrology. In the linear setting, near-Heisenberg scaling and enhanced robustness against phase damping are observed for suitably tailored Dicke superposition probes. In the nonlinear setting, optimal and near-optimal Dicke superposition probes are identified for representative two-body interaction Hamiltonians;  their performance is systematically analyzed under local and global noise channels. For the system sizes considered, near-optimal Dicke superposition probes retain a substantial metrological advantage over the nonlinear shot-noise limit in the weak-noise regime. These results establish Dicke superposition states as promising multipartite resources for quantum-enhanced sensing.

\emph{Note added:} After completion of the original version of this manuscript, we became aware of the recent work of Li and Ren~\cite{LiRen2026}, where the Dicke superposition states $\lvert D^{(N)}_{l,l'}\rangle$ specified in Eqs.~(\ref{odd}) and (\ref{even}) were identified as near-optimal probe states in the noiseless linear-encoding scenario. The results on Dicke superposition probes presented in this work were obtained independently. A brief comparison with Ref.~\cite{LiRen2026} has been included in the concluding paragraph of Sec.~III.

\noindent {\bf Acknowledgments}: Sudha and BNK acknowledge the financial support of KITS, K-Tech, Karnataka Govt under the Q-Pragathi project.
\bibliographystyle{apsrev4-2}
\bibliography{referencesR}
\end{document}